\documentclass[12pt]{article}
\usepackage{amssymb}
\usepackage{amsfonts}
\usepackage{xcolor}
\usepackage{color}
\usepackage{amsmath}
\usepackage[normalem]{ulem}
\usepackage[margin = 1in]{geometry}
\usepackage{natbib}
\usepackage[doublespacing]{setspace}
\usepackage{amsthm}
\usepackage[flushleft]{threeparttable}
\usepackage{adjustbox}
\usepackage{caption}
\usepackage{subcaption}
\usepackage{float}
\usepackage{array}
\usepackage{dsfont}

\usepackage{filecontents}

\usepackage{hyperref}
\hypersetup{
    colorlinks=true,
    linkcolor=black,     
    urlcolor=black,
    citecolor=black,
    }

\newtheorem{theorem}{Theorem}

\newtheorem{assumption}{Assumption}[section]

\newtheorem*{corollary}{Corollary}

\newtheorem{lemma}{Lemma}[section]

\newtheorem{proposition}[theorem]{Proposition}

\theoremstyle{definition}

\newtheorem{remark}{Remark}
\newtheorem*{remark*}{Remark}

\makeatletter
\newcommand*{\rom}[1]{\expandafter\@slowromancap\romannumeral #1@}
\makeatother

\date{October 2025}

\title{ \textbf{Testing for Peer Effects without Specifying the Network Structure}\thanks{%
We thank the Editor {\'A}ureo de Paula, the Associate Editor and three anonymous referees for their valuable comments and suggestions. We are also grateful to Stanislav Anatolyev, John Chao, Patrik Guggenberger, Michael P. Leung, Adam McCloskey, Chris Muris, Ryo Okui, Wendun Wang, Yulong Wang, and the participants of the seminars at Syracuse, Osaka, Tokyo, the 28th International Panel Data Conference, the 2023 Meeting of the Midwest Econometrics Group and the XVIII World Conference of Spatial Econometrics Association for helpful comments and discussions at various stages of the project.} }

\author{Hyunseok Jung \qquad Xiaodong Liu\thanks{Jung: Department of Economics, University of Arkansas, Fayetteville, Arkansas 72703, USA, hj020@uark.edu. Liu: Corresponding author. Department of Economics, University of Colorado Boulder, Boulder, Colorado 80309, USA, xiaodong.liu@colorado.edu. 
}}

\begin{document}

\maketitle
\newpage
		\begin{abstract}
        \doublespacing
		This paper proposes an Anderson-Rubin (AR) test for the presence of peer effects in panel data without the need to specify the network structure. The unrestricted model of our test is a linear panel data model of social interactions with dyad-specific peer effect coefficients for all potential peers. The proposed AR test evaluates if these peer effect coefficients are all zero. As the number of peer effect coefficients increases with the sample size, so does the number of instrumental variables (IVs) employed to test the restrictions under the null, rendering a many-IV environment of \citet{Bekker1994}.         By extending existing many-IV asymptotic results to panel data, we establish the asymptotic validity of the proposed AR test. Our Monte Carlo simulations show the robustness and improved  
        performance of the proposed test compared to some existing tests with misspecified networks. We provide two applications to demonstrate its empirical relevance.\\
		
  
\end{abstract}
\newpage

\section{Introduction}
A major stumbling block in the study of network effects is the need to specify the interaction structure. Most existing estimators and tests for network effects require \emph{a priori} specification of the underlying network. 
Researchers often use data, if available, on geographical, economic, or social relationships between individuals (e.g., bilateral trade volume, friendship survey, etc.), along with a set of user-chosen rules (e.g., inverse distance, k-nearest neighbors, etc.), to determine the presence and strength of network connections in their models. However, the resulting 
network structure is subject to potential misspecification, and the misspecification risk is more severe when the specified network structure is purely based on theories or assumptions due to data limitations (e.g., linear-in-means).

A new body of literature on the identification and testing of network effects has emerged to tackle this issue. \citet{Blume2015} show that identification of network effects is possible even if the network structure is only partially known, as long as there are two individuals who are \emph{a priori} known to be unconnected. \citet{Breza2020} propose a technique to estimate social links using aggregated relational data.
\citet{BPR2021} introduce a new equilibrium concept for network formation models called ``network competitive equilibrium," which allows recovery of unobserved social networks using only observable outcomes. \citet{LQT23} propose an identification strategy for cross-sectional social interaction models with many small networks, where unobserved network links are treated as random variables and network effects are identified from the ``mean" relationship between the reduced form coefficients and structural parameters. 
For panel data models, several papers exploit the sparsity of network links commonly observed in social networks 
to directly estimate individual links using shrinkage estimation methods \citep[e.g.,][]{Bonaldi2015, Manresa2016, Rose2018}. More recently, \citet{DRS24} consider a panel data model similar to those in the aforementioned papers, but their identification relies on differential popularity across individuals in a network, instead of the sparsity assumption. 

To test for network effects, 
\citet{LP2018, LP2025} extend the Moran I test \citep{Moran} to accommodate situations where the researcher faces multiple possible specifications of the underlying network structure. In the spatial econometric literature, some papers \citep[e.g.,][among others]{Ng2006, PUY2008, SYR2009, BFK2012, CGL2012, Pesaran2021} consider tests for cross-sectional correlation in panel data models with unspecified correlation structure. However, these tests are primarily designed for detecting cross-sectional correlation in the error term.

We contribute to this fast-growing literature by proposing an Anderson-Rubin (AR) test for the presence of peer effects that does not require specifying the network structure. The unrestricted model of our test is a linear panel data model of social interactions with dyad-specific peer effect coefficients for all potential peers. When no information on the network structure is available, all the other individuals in the network can be treated as potential peers. Our AR test evaluates if these peer effect coefficients are all zero.

Our test does not require the estimation of individual network links or peer effect coefficients, so it does not require restrictive regularity conditions and is much easier to implement than most existing methods in the literature. 
However, the merit comes at the cost of not being able to identify the strength of peer effects. Therefore, our test can be especially useful when the \textit{presence} of peer effects is the primary concern or interest.\footnote{{Our AR test does not distinguish between \emph{endogenous} and \emph{exogenous} peer effects \citep{Manski93} because these two types of effects cannot be disentangled without any information or restrictions on
the network \citep[see Theorems 2 and 6 of][and Remark \ref{rmk3} of this paper]{Blume2015}.} Therefore, the main objective of our test is to detect any forms of peer effects, rather than to identify the exact nature of the effects.} For instance, suppose a researcher conducts a causal analysis that relies on the Stable Unit Treatment Value Assumption, which requires no spillovers between treated and control units.
To provide supportive evidence for this crucial assumption, the researcher may report some test statistics along with the estimation results. The proposed AR test is particularly advantageous in this context as it does not require costly and time-consuming data collection on network links or estimation of these links.
On the other hand, if the presence of peer effects is the primary interest, our test can provide general evidence for peer effects that is not contingent on any specific assumption regarding the network structure, as demonstrated in our empirical applications.\footnote{A common strategy in applied research for identifying spatial or social effects is to use the spatial lags of covariates  
as instrumental variables (IVs) for endogenous spillover  
effects. 
However, as discussed in \citet{GO2012}, this approach is vulnerable to weak identification. 
For instance, when the interaction structure is misspecified, the IVs derived based on 
the specified network may become invalid. Even when the network structure is correctly specified, the spatial lags of covariates are often highly correlated with each other or with other terms in the model, resulting in insufficient variation to identify the spillover  
effects. 
In such cases, 
our test can be an effective alternative to the existing identification strategy, as it does not rely on 
any parametric 
assumptions about the network structure.}

Our test is also closely related to the literature on inference with many instruments and/or many restrictions \citep[e.g.,][among others]{Bekker1994, Donald2003, AG2011, LO2012, Chao2014, Crudu2021, MS2022, AS2023}. In our test, as the number of dyad-specific peer effect coefficients increases with the sample size, so do the number of restrictions under the null and the number of instrumental variables (IVs) employed to test the restrictions, leading to 
a testing problem with many restrictions and many IVs. To find a sufficient number of IVs to test the restrictions under the null, we exploit the exogenous characteristics of potential peers.\footnote{Our choice of IVs follows \cite{DRS24}, but their approach assumes the number of agents in the network is fixed so that the number of IVs is fixed, which is different from our many-IV setting.} This is a unique many-IV scenario that arises naturally in the inference of network models without information on the network structure, and the two burgeoning research areas are nicely connected in this paper. 

In this paper, we first illustrate the main idea of the proposed test using a simple panel data model without fixed effects. Then, we extend the test to a panel data model that includes both individual and time fixed effects.
To the best of our knowledge, our paper is among the first to analyze \citet{Bekker1994}'s many-IV problem in a panel data model with two-way fixed effects.
By adapting existing many-IV asymptotic results \citep[e.g.,][]{Hansen2008, AG2011, Chao2012, MS2022} to the panel data setting, we show that, under the null, our test statistic is asymptotically normal and has the correct size, 
allowing the number of agents in the network to increase to infinity at the same rate as the number of time periods.\footnote{The asymptotic validity of our test requires $nT \to \infty$ and $n < T$, where $n$ denotes the number of individuals and $T$ the number of time periods. So, our test requires a long panel. 
On the other hand, the power analysis in Remark \ref{power} suggests that when the number of null restrictions violated (i.e., the number of nonzero dyadic-specific peer effect coefficients) is fixed, consistency of our test requires that $T$ grows faster than $n$. Nevertheless, our Monte Carlo simulations indicate that the AR test performs reasonably well even when $n$ is comparable to $T$, such as $(n,T) = (30,50)$ or $(40,50)$. With the increasing availability of long panels in empirical research, our test is applicable in a wide range of settings.}

We conduct Monte Carlo simulations to investigate the finite sample performance of the proposed AR test. To study the power properties of the test, we consider various levels of network sparsity and interaction intensity. Furthermore, our simulations show the robustness and improved performance of the proposed AR test compared to some existing tests when the network is misspecified.
We also provide two empirical applications to demonstrate how the proposed AR test can be applied in practice.

The remainder of the paper is organized as follows: Section \ref{ARwof} and \ref{ARwf} introduce the models and test statistics in the absence and presence of fixed effects; Section \ref{simul} conducts Monte Carlo simulations examining empirical size and power of the test; Section \ref{emp} applies the AR test to two empirical models: international growth spillover and National Basketball Association (NBA) player interaction models; and Section \ref{con} concludes. All the proofs of the theoretical results in this paper, the full version of the power analysis presented in Remark \ref{power}, additional simulation results, and the estimator for the excess kurtosis of regression error discussed in Section \ref{ARwf} are contained in Appendix. 

Throughout this paper, we follow the convention of using boldface uppercase letters for matrices and row vectors, and boldface lowercase letters for column vectors.

\section{AR Test for Peer Effects}\label{ARwof}
In this section, we illustrate the main idea of the proposed test using a simple panel data model without fixed effects. We consider a more general panel data model with both individual and time fixed effects in Section \ref{ARwf}.

Consider a set of $n$ individuals $\mathcal{N}=\{1,2,\cdots ,n\}$. Let $%
\mathcal{N}_{i}$ denote the set of potential peers of individual $i$ and $%
n_{i}\equiv |\mathcal{N}_{i}|$ denote the cardinality of $\mathcal{N}_{i}$. When no information on $\mathcal{N}_{i}$ is available, all the other individuals in $\mathcal{N}$ can be treated as potential peers of individual $i$, i.e., $\mathcal{N}_{i}=\mathcal{N}/\{i\}$. In certain situations, researchers may know \emph{a priori} that network links do not exist between certain pairs of individuals.\footnote{For instance, researchers may have prior knowledge that spillovers do not occur between some predetermined groups or clusters of individuals (see Remark \ref{rmk5} for further discussion).
We thank the Associate Editor for raising this point.} This information will reduce the number of potential peers of some individuals and thus the number of restrictions under the null hypothesis as we will see below. However, our test in general does not require any knowledge of $\mathcal{N}_{i}$, and this scenario is the primary focus of the paper. 
Suppose the outcome of individual $i$ in period $t$ is given by%
\begin{equation}
y_{it}=\sum_{j\in \mathcal{N}_{i}}\alpha _{ij}y_{jt}+\mathbf{X}_{it}\boldsymbol{\beta} +u_{it},
\label{model1}
\end{equation}%
for $i=1,\cdots ,n$ and $t=1,\cdots ,T$, where $\mathbf{X}_{it}$ is a $L$-dimensional row-vector of exogenous variables and $u_{it}$ is the error term. The coefficients $\alpha _{ij}$ represent dyad-specific \emph{endogenous} peer effects \citep{Manski93}.\footnote{In Remark \ref{rmk3}, we point out that a significant value of our test statistic indicates the presence of either \emph{endogenous} or \emph{exogenous} peer effects \citep{Manski93}. 
See Remark \ref{rmk3} for further discussion.}
Our goal is to test for the presence of peer effects, i.e., $%
H_{0}:\alpha _{ij}=0$ for all potential pairs of peers $(i,j)$.\footnote{An alternative approach to testing this joint null hypothesis is to test each restriction $\alpha _{ij}=0$
individually, using a Bonferroni-type procedure to control the familywise error rate (FWER).
However, the Bonferroni procedure is known to be conservative, especially when the number of hypotheses is large, as it does not account for the dependence structure among the test statistics associated with each hypothesis. Therefore it may be desirable to test the restrictions jointly, as we propose in this paper. In addition, asymptotically controlling the FWER becomes nontrivial when the number of hypotheses increases with the sample size.
For these reasons, we leave this direction for future research. 
We thank an anonymous referee for raising this point.}
As the number of peer effect coefficients $\alpha _{ij}$ is proportional to the number of potential dyads in the network, the null hypothesis of our test imposes many restrictions \citep[see][for recent developments on testing with many restrictions]{AS2023}.\footnote{The magnitude and density (i.e., the number of non-zero $\alpha_{ij}$) of the peer effect coefficients determine how strongly individuals are connected in the network and thus characterize the degree of deviation from the null hypothesis. A sparse network (or one with a small number of non-zero $\alpha_{ij}$) corresponds to a \emph{weak violation} of the null. When the alternative is only weakly separated from the null, all tests inherently exhibit low power. In Remark \ref{power}, we provide analytical results on how network density affects the power of our test.}

The peer effect term can be written more compactly as $\sum_{j\in \mathcal{N}%
_{i}}\alpha _{ij}y_{jt}=\mathbf{Y}_{it}\boldsymbol{\alpha} _{i}$, where $\mathbf{Y}_{it}$ is a row vector
collecting the outcomes of individual $i$'s potential peers and $\boldsymbol{\alpha _{i}}$ is a column vector of corresponding coefficients. Let $\mathbf{e}_{i}$ denote the $i$th
column of the identity matrix $\mathbf{I}_{n}$ and $\boldsymbol{\iota} _{n}$ denote an $n\times 1$
vector of ones. In matrix form, Equation (\ref{model1}) can be written as%
\begin{equation*}
\mathbf{y}_{t}=\mathbf{Y}_{t}\boldsymbol{\alpha} +\mathbf{X}_{t}\boldsymbol{\beta} +\mathbf{u}_{t},
\end{equation*}%
for $t=1,\cdots ,T$, where $\mathbf{y}_{t}=(y_{1t},\cdots ,y_{nt})^{\prime }$, $%
\mathbf{Y}_{t}=(\mathbf{e}_{1}\mathbf{Y}_{1t},\cdots ,\mathbf{e}_{n}\mathbf{Y}_{nt})$, $\boldsymbol{\alpha} =(\boldsymbol{\alpha} _{1}^{\prime
},\cdots ,\boldsymbol{\alpha} _{n}^{\prime })^{\prime }$, $\mathbf{X}_{t}=(\mathbf{X}_{1t}^{\prime },\cdots
,\mathbf{X}_{nt}^{\prime })^{\prime }$, and $\mathbf{u}_{t}=(u_{1t},\cdots ,u_{nt})^{\prime }$%
. Stacking the observations over the $T$ periods together, we have%
\begin{equation}
\mathbf{y}=\mathbf{Y}\boldsymbol{\alpha} +\mathbf{X}\boldsymbol{\beta} +\mathbf{u},  \label{model1_vec}
\end{equation}%
where $\mathbf{y}=(\mathbf{y}_{1}^{\prime },\cdots ,\mathbf{y}_{T}^{\prime })^{\prime }$, $%
\mathbf{Y}=(\mathbf{Y}_{1}^{\prime },\cdots ,\mathbf{Y}_{T}^{\prime })^{\prime }$, $\mathbf{X}=(\mathbf{X}_{1}^{\prime
},\cdots ,\mathbf{X}_{T}^{\prime })^{\prime }$, and $\mathbf{u}=(\mathbf{u}_{1}^{\prime },\cdots
,\mathbf{u}_{T}^{\prime })^{\prime }$.

\begin{remark} [SAR models]\label{rmk1}
The model defined in Equation (\ref{model1_vec}) reduces to a standard spatial autoregressive (SAR) panel data model 
\begin{equation*}
\mathbf{y}=\rho (\mathbf{I} _{T}\otimes\mathbf{W})\mathbf{y}+\mathbf{X}\boldsymbol{\beta} +\mathbf{u},
\end{equation*}%
if $\alpha _{ij}=\rho w_{ij}$, where $\rho$ represents the \emph{homogeneous} peer effect and $w_{ij}$ is the $(i,j)$th element of the $n\times n$ adjacency matrix $\mathbf{W}$. 
To test $H_{0}:\rho=0$, it is often assumed that the underlying network structure captured by the adjacency matrix $\mathbf{W}$ is known and exogenously predetermined, and the row and column sums of the matrices $\mathbf{W}$ and $[\mathbf{I} _N-\rho (\mathbf{I} _{T}\otimes\mathbf{W})]^{-1}$ are bounded uniformly in absolute value.\footnote{Recently, \cite{PY21} and \cite{LYY23} develop central limit theorems allowing for some columns of the adjacency matrix $\mathbf{W}$ to have unbounded sums.}
Some of these assumptions are hard to verify \emph{a priori}. 
In contrast, our AR test does not rely on any specific network structure to detect peer effects, and thus does not require any of these restrictive assumptions. Our Monte Carlo simulations in Section \ref{simul} show the robustness and improved performance of the AR test compared to some existing tests based on SAR models when the assumptions are violated. $\blacksquare$
\end{remark}

The outcomes of individual $i$'s potential peers contained in $\mathbf{Y}_{it}$ are endogenous
and natural instruments for $\mathbf{Y}_{it}$ are the exogenous characteristics 
of individual $i$'s potential peers denoted by $\mathbf{Z}_{it}$ \citep[see, e.g.,][]{DRS24}. For instance, if $\mathcal{N}_{i}=\mathcal{N}/\{i\}$, then we could
use $\mathbf{Z}_{it}=[\mathbf{X}_{jt}]_{j\in \mathcal{N}_{i}}=(\mathbf{X}_{1t},\cdots
,\mathbf{X}_{i-1,t},\mathbf{X}_{i+1,t},\cdots ,\mathbf{X}_{nt})$ as instruments for $%
\mathbf{Y}_{it}=(y_{1t},\cdots ,y_{i-1,t},y_{i+1,t},\cdots ,y_{nt})$.\footnote{\label{ft2}
In general, we could use linear combinations of the potential peers' exogenous characteristics as IVs for their outcomes. For instance, when $%
\mathcal{N}_{i}=\mathcal{N}/\{i\}$, we could use $\mathbf{Z}_{it}=(\mathbf{X}_{1t}\mathbf{B},\cdots
,\mathbf{X}_{i-1,t}\mathbf{B},\mathbf{X}_{i+1,t}\mathbf{B},\cdots ,\mathbf{X}_{nt}\mathbf{B})$ as IVs for $%
\mathbf{Y}_{it}=(y_{1t},\cdots ,y_{i-1,t},y_{i+1,t},\cdots ,y_{nt})$, where $\mathbf{B}$ is a $%
L\times q$ matrix of known constants. If $\mathbf{B}=\mathbf{I}_{L}$, then $%
\mathbf{Z}_{it}=(\mathbf{X}_{1t},\cdots ,\mathbf{X}_{i-1,t},\mathbf{X}_{i+1,t},\cdots ,\mathbf{X}_{nt})$. If $L$, the
dimension of $\mathbf{X}_{it}$, is large, then we could choose a $\mathbf{B}$ with a small $q$ to
reduce the number of IVs in $\mathbf{Z}$.} Let $\mathbf{Q}$ denote the IV matrix
collecting linearly independent columns in $[\mathbf{X},\mathbf{Z}]$, where $\mathbf{Z}=(\mathbf{Z}_{1}^{\prime
},\cdots ,\mathbf{Z}_{T}^{\prime })^{\prime }$ with $\mathbf{Z}_{t}=(\mathbf{e}_{1}\mathbf{Z}_{1t},\cdots
,\mathbf{e}_{n}\mathbf{Z}_{nt})$, and $K$ denote the number of columns in $\mathbf{Q}$. If the number
of potential peers increases with $n$ such that $n_{i}=O(n)$ (e.g., when $\mathcal{N}_{i}=\mathcal{N}/\{i\}$), 
then the
dimension of $\mathbf{Z}_{it}$ is $O(n)$ and, hence, $K=O(n^{2})$. 
Therefore, we are in the many-IV framework of \citet{Bekker1994}.

Let $N=nT$, $\mathbf{P}=\mathbf{Q}(\mathbf{Q}^{\prime }\mathbf{Q})^{-1}\mathbf{Q}^{\prime }$ and $\mathbf{D}$ be a diagonal matrix
containing the diagonal elements of $\mathbf{P}$. Using residuals computed from the restricted model $\widetilde{\mathbf{u}}=\mathbf{y}-\mathbf{X}\widetilde{%
\boldsymbol{\beta} }$, where $\widetilde{\boldsymbol{\beta} }=(\mathbf{X}^{\prime }\mathbf{X})^{-1}\mathbf{X}^{\prime }\mathbf{y}$, we adopt 
the following jackknife AR test statistic \citep[e.g.,][]{MS2022} for $H_{0}:\boldsymbol{\alpha} =\mathbf{0}$%
\begin{equation}
AR=\frac{1}{\sqrt{K}\sqrt{\widetilde{\Phi }}}\widetilde{\mathbf{u}}^{\prime }(\mathbf{P}-\mathbf{D})%
\widetilde{\mathbf{u}},  \label{AR1}
\end{equation}%
where $\widetilde{\Phi }$ is a consistent estimator of $\Phi =\mathrm{plim}%
_{N\rightarrow \infty }\frac{2}{K}\mathrm{tr}[(\mathbf{P}-\mathbf{D})\boldsymbol{\Omega} (\mathbf{P}-\mathbf{D})\boldsymbol{\Omega} ]$,
with $\boldsymbol{\Omega} =\mathrm{E}(\mathbf{u}\mathbf{u}^{\prime }|\mathbf{X})$. The jackknife AR test statistic defined in Equation (\ref{AR1}) removes  
$\sum_{i}p_{ii}\widetilde{u}_{i}^2$ (= $\widetilde{\mathbf{u}}^{\prime }\mathbf{D}\widetilde{\mathbf{u}}$), where $p_{ii}$ denotes the $i$th diagonal element of $\mathbf{P}$, from the quadratic form $\widetilde{\mathbf{u}}^{\prime }\mathbf{P}\widetilde{\mathbf{u}}$ to re-center the test statistic to zero.  
This re-centering is referred to as leave-one-out or jackknife in the literature. The main advantage of the jackknife method is its robustness to
heteroskedasticity of unknown form. This 
idea was introduced by \cite{Angrist1999} and has generated a rich literature in econometrics \cite[e.g.,][]{Hausman2012,Chao2012,Bekker2015,Crudu2021,MS2022}.

\citet{Crudu2021} suggest 
estimating $\Phi $ by $\widetilde{\Phi }=\frac{2}{K}\mathrm{tr}[(\mathbf{P}-\mathbf{D})%
\widetilde{\boldsymbol{\Omega} }(\mathbf{P}-\mathbf{D})\widetilde{\boldsymbol{\Omega} }]$, where $\widetilde{\boldsymbol{\Omega} }=%
\mathrm{diag}\{\widetilde{u}_{it}^{2}\}$.
\citet{MS2022} point out that this variance estimator can lead to a loss of power 
and propose an alternative estimator for $%
\Phi $. For brevity,
we refer interested readers to thorough discussions on the consistent
estimation of $\Phi $ in \citet{MS2022}.

In our approach, for each endogenous regressor, $y_{jt}$, in Equation (\ref{model1}) (i.e., the outcome of potential peer $j$), we use its corresponding exogenous characteristics, $\mathbf{X}_{jt}$, as the IVs. In other studies analyzing AR tests with many ``weak'' instruments \citep[e.g,][]{MS2022},  IVs are introduced through an auxiliary model (i.e., a first-stage regression) and are often weak in practice due to the difficulty of finding exogenous variables that affect the outcome $y$ only through the endogenous regressor. In contrast, in our setting, IVs are implicitly defined within the main regression model, i.e., Equation (\ref{model1}), and thus are strong as long as some of the exogenous characteristics $\mathbf{X}$ are informative in predicting the outcomes. They remain strong even under the null hypothesis of no spillover effects. Since most applications include at least some exogenous variables that are predictive, we do not consider weak identification as a major concern for our test.

Another important feature of our testing environment, compared to existing studies on testing with many IVs, is that the number of parameters of interest or the number of restrictions under the null diverges with $n$ (i.e., testing with many IVs \& many restrictions). The literature has typically studied the two testing problems separately -- testing with many IVs \citep[e.g.,][]{Bekker1994,Donald2003,AG2011,LO2012,Chao2014,Crudu2021,MS2022} and testing with many restrictions \citep[e.g.,][]{ANA2012,CAL2011,Catt2018a,Catt2018,AS2019,AS2023}. However, as \citet{AS2019} notes, the asymptotic tools used for the two problems are often the same or similar, as they result in comparable test statistics (e.g., bilinear forms) and require similar econometric treatments to handle increasing dimensionality. Furthermore, our AR test statistic is evaluated under the null, where all dyad-specific peer effect coefficients are restricted to zero. As a result, the presence of many parameters in the unrestricted model does not materially affect the test statistic or its asymptotic behavior under the null. This is clearly an important merit of using the AR test in our setting, as it allows us to circumvent complications associated with estimating a large number of parameters \citep[e.g.,][]{AS2023}.

To establish the asymptotic validity of the test statistic defined in Equation
(\ref{AR1}), we maintain the following assumptions.

\begin{description}
\item[Assumption 1] The errors $u_{it}$ are independent across $i$ and $t$,
with $\mathrm{E}(u_{it}|\mathbf{X}_{it})=0$, $\mathrm{E}(u_{it}^{2}|\mathbf{X}_{it})=\varsigma
_{it}^{2}\geq \underline{\varsigma }^{2}$, for some constant $\underline{%
\varsigma }^{2}>0$, and uniformly bounded fourth conditional moments.

\item[Assumption 2] The IV matrix $\mathbf{Q}$ has full column rank $K$, $K\rightarrow
\infty $ as $N\rightarrow \infty $, and there exists a constant $C_{p}$ such
that $p_{ii}\leq C_{p}<1$, where $p_{ii}$ is the $i$th diagonal element of $%
\mathbf{P}=\mathbf{Q}(\mathbf{Q}^{\prime }\mathbf{Q})^{-1}\mathbf{Q}^{\prime }$.

\item[Assumption 3] $\mathrm{plim}_{N\rightarrow \infty }\frac{1}{N}%
\mathbf{X}^{\prime }\mathbf{X}$ is finite and nonsingular. $\mathrm{plim}_{N\rightarrow \infty
}\frac{1}{N}\mathbf{X}^{\prime }\mathbf{DX}$, $\mathrm{plim}_{N\rightarrow \infty }\frac{1}{N}%
\mathbf{X}^{\prime }\boldsymbol{\Omega}\mathbf{X}$, and $\mathrm{plim}_{N\rightarrow \infty }\frac{1}{N}%
\mathbf{X}^{\prime }\mathbf{D}\boldsymbol{\Omega} \mathbf{DX}$ are finite.
\end{description}

In the literature on inference with many instruments and/or many restrictions, it is common to assume that the error terms are independent \citep[e.g.,][]{Donald2003, AG2011, LO2012, Chao2014, Crudu2021, MS2022, AS2023}. 
In the next section, we extend the model by introducing two-way fixed effects to partially account for dependence across individuals and over time.
Assumption 2 implies that $\frac{1}{N}\mathrm{tr}(\mathbf{P})=\frac{K}{N}\leq C_{p}<1$. 
In our setting, as $K=O(n^{2})$ and $N=nT$, Assumption 2 allows $n$ to grow at the same rate as $T$, subject to the condition $n<T$. 
The main advantage of the AR test is that it only requires estimation of the restricted
model, and
Assumption 3 ensures that the exogenous variables $\mathbf{X}$ have enough variation 
and the OLS estimator for the restricted model 
is well behaved.
The following proposition establishes the asymptotic
normality of the proposed test statistic under the null hypothesis.\footnote{When $n$ is fixed, the number of IVs is fixed. Then, the asymptotic distribution of our test statistic reduces to a chi-square distribution with $K$ degrees of freedom (see the corollary to Proposition \ref{prop2} in Section \ref{ARwf} for more details).}

\begin{proposition}
\label{prop1}Suppose Assumptions 1-3 hold and $\widetilde{\Phi }$ is a
consistent estimator of $\Phi $. Under $H_{0}:\boldsymbol{\alpha} =\mathbf{0}$, the AR test
statistic defined in Equation (\ref{AR1}) is asymptotically standard normal.
\end{proposition}

\begin{remark} [Power analysis] \label{power}

To gain insight into how key parameters in the unrestricted model, such as the peer effect coefficients, affect 
the power of our test, we derive a power formula under some simplifying assumptions in what follows.

We consider an alternative, where the first $m$ out of $n$ individuals have at least one incoming or outgoing link and the rest are isolated from the network. The size of $m$ relative to $n$ reflects the density of the underlying network. Let $\mathbf{A}_n=[\alpha_{ij}]$ denote the $n \times n$
network adjacency matrix, where non-zero elements are confined to the upper-left $m \times m$ submatrix. Then, under the alternative, the outcome model in period $t$ can be written as%
\begin{equation*} 
\mathbf{y}_t=\mathbf{A}_n\mathbf{y}_t+\beta\mathbf{x}_t + \mathbf{u}_t,
\end{equation*}%
where, for simplicity, we assume that there is a single exogenous regressor $\mathbf{x}_t$ in the model and the coefficient $\beta$ is known.\footnote{In Appendix \ref{App_Power_Analysis}, we discuss how the estimation of $\beta$ affects the power of the test.} These simplifications do not alter the main message of the analysis.

In Appendix \ref{App_Power_Analysis}, we show that the AR test statistic can be decomposed into three components: one deterministic and two stochastic, and for the test to be consistent, the deterministic component must diverge, requiring
\begin{eqnarray} \label{con1}
 \frac{\beta^2 (1-C_p)}{\sqrt{K\Phi }} \sum_{t=1}^{T} \sum_{i=1}^{m}\left(\sum_{j=1}^{m} g_{ij}x_{jt} \right)^2 \to \infty, 
\end{eqnarray}
where $C_p < 1$ is the upper bound of the diagonal elements of $\mathbf{P}$ defined in Assumption 2 and $g_{ij}$ denotes the $(i,j)$th entry of $\mathbf{A}_n(\mathbf{I}_{n}-\mathbf{A}_n)^{-1}$. As the reduced form of the alternative model in Appendix \ref{App_Power_Analysis} implies, the term $\beta^2\sum_{t=1}^{T} \sum_{i=1}^{m}\left(\sum_{j=1}^{m} g_{ij}x_{jt} \right)^2$ quantifies the strength of peer effects in the data and thus determines the power of our test. We refer to (\ref{con1}) as the power formula. 

Given $K = O(n^2)$ in this setup and under the maintained assumption $x_{it} = O_p(1)$, we use the power formula to evaluate the power of our AR test for several representative network structures. 
First, suppose that $m$ is fixed and $\sum_{j=1}^{m} g_{ij}$ is bounded. For instance, $\alpha_{12}$ is a nonzero constant and all the other elements of $\mathbf{A}$ are zero. Then, $\sum_{i=1}^{m}\left(\sum_{j=1}^{m} g_{ij}x_{jt} \right)^2$ is bounded and consequently, according to (\ref{con1}), consistency requires that the number of time periods $T$ grows faster than the number of individuals $n$. This result is intuitive: when $m$ is fixed, the overall level of peer effects in the network becomes increasingly diluted as $n$ grows. Therefore, a larger time dimension is necessary to accumulate sufficient information 
to detect the peer effects.

Next, consider a network where every individual has a bounded number of connections. An example is the nearest neighbor design considered in our simulation study. In this case, $\sum_{j=1}^{m}g_{ij}$ is bounded, but $m$, the number of individuals with network connections, increases with $n$. As a result, the test can be consistent even when $T$ increases at the same rate as $n$. In some networks, there exist individuals with an unbounded number of outgoing links (i.e., dominant units). As argued by \citet{PY2021}, the presence of dominant units is a common feature of real-world networks. In this setting, $m$ increases with $n$ and thus the test is consistent even when $T$ increases at the same rate as $n$.

In the network literature, a network is considered dense when the number of links is of order $n^2$.
An example is a random graph where every pair of nodes has a fixed nonzero probability of forming a link. Clearly, in this case, $\sum_{i=1}^{m}\left(\sum_{j=1}^{m} g_{ij}x_{jt} \right)^2$ increases with $n$ and the test is consistent even when $T$ increases at the same rate as $n$.

In summary, if the number of null restrictions violated (i.e., instances where $\alpha_{ij} \neq 0$) is fixed, 
consistency requires the time dimension $T$ to grow faster than the cross-sectional dimension $n$. In contrast, if the number of these violations
increases with $n$, consistency can be achieved even when $n$ and $T$ grow at the same rate. Our Monte Carlo simulations in Section \ref{simul} examine how these parameters affect the power of the test in finite samples.\footnote{Also, the power formula indicates that the power depends on $\boldsymbol{\beta}$, the coefficients of the exogenous characteristics. These coefficients measure the predictive strength of the characteristics and, consequently, the strength of IVs, which in turn affects the power of the test.} $\blacksquare$
\end{remark}

\begin{remark} [Testing for exogenous peer effects] \label{rmk3}
The proposed AR test is based on the exogeneity condition $\mathrm{E}(\mathbf{Q}^{\prime }\widetilde{\mathbf{u}})=\mathbf{0}$,
where $\mathbf{Q}$ is the IV matrix collecting linearly independent columns in $[\mathbf{X},\mathbf{Z}]$ and $\widetilde{\mathbf{u}}$ is the residual vector computed from the restricted model.
Hence, the AR test statistic for $H_0:\boldsymbol{\alpha} = \mathbf{0}$ in Equation (\ref{model1_vec}) is numerically identical to that for $H_0:\boldsymbol{\gamma} = \mathbf{0}$ in 
\begin{equation}
\mathbf{y}= \mathbf{X}\boldsymbol{\beta} +\mathbf{Z}\boldsymbol{\gamma} +\mathbf{u}.  
\label{model2_vec}
\end{equation}%
This issue is, in spirit, similar to that encountered when testing for over-identifying restrictions, where a significant test statistic suggests either that the instruments are asymptotically correlated with the error terms or that some of the instruments have been incorrectly omitted from the regression equation.

Recall $\mathbf{Z}=(\mathbf{Z}_{1}^{\prime},\cdots ,\mathbf{Z}_{T}^{\prime })^{\prime }$, where $\mathbf{Z}_{t}=(\mathbf{e}_{1}\mathbf{Z}_{1t},\cdots,\mathbf{e}_{n}\mathbf{Z}_{nt})$
and $\mathbf{Z}_{it}=[\mathbf{X}_{jt}]_{j\in \mathcal{N}_{i}}$ is a row vector containing exogenous characteristics of individual $i$'s potential peers $j\in \mathcal{N}_{i}$. Then, Equation (\ref{model2_vec}) can be rewritten as 
\begin{equation}
y_{it}=\mathbf{X}_{it}\boldsymbol{\beta} +\sum_{j\in \mathcal{N}_{i}}\mathbf{X}_{jt}\boldsymbol{\gamma} _{ij}+u_{it},
\label{model3}
\end{equation}%
where $\boldsymbol{\gamma} _{ij}$ captures the influence of potential peers' exogenous characteristics and can be viewed as dyad-specific \emph{exogenous} peer effects or \emph{contextual} effects \citep{Manski93}. 
Therefore, a significant value of the proposed AR test statistic indicates the presence of either \emph{endogenous} peer effects, i.e., $\alpha_{ij}\neq 0$ in Equation (\ref{model1}), or \emph{exogenous} peer effects, i.e., $\boldsymbol{\gamma} _{ij}\neq \mathbf{0}$ in Equation (\ref{model3}). Intuitively, this follows because peers’ characteristics are used to evaluate the null hypothesis in both cases -- either as IVs for the endogenous peer effects in Equation (\ref{model1}) or as potential covariates in Equation (\ref{model3}). 

This is related to the fundamental identification issue discussed in  \citet[][Theorems 2 and 6]{Blume2015} that, without knowing the network structure or how individuals interact, 
these two types of peer effects cannot be disentangled. More specifically, a model that includes both endogenous and exogenous peer effects
\begin{equation}
y_{it}=\sum_{j \in \mathcal{N}/\{i\}}\alpha _{ij}y_{jt}+\mathbf{X}_{it}\boldsymbol{\beta} +\sum_{j \in \mathcal{N}/\{i\}}\mathbf{X}_{jt}\boldsymbol{\gamma} _{ij} +u_{it}
\label{model-mix1}
\end{equation}%
is generally not identifiable without imposing additional restrictions \citep[see, e.g.,][]{DRS24}.\footnote{This identification problem can be easily seen in the simple case with $n=2$. In this case, the model with both endogenous and exogenous peer effects is given by%
\begin{eqnarray*}
y_{1t} &=&\alpha _{12}y_{2t}+\mathbf{X}_{1t}\boldsymbol{\beta} +\mathbf{X}_{2t}\boldsymbol{\gamma}_{12}+u_{1t} \\
y_{2t} &=&\alpha_{21}y_{1t}+\mathbf{X}_{2t}\boldsymbol{\beta} +\mathbf{X}_{1t}\boldsymbol{\gamma}_{21}+u_{2t},
\end{eqnarray*}%
where $\alpha _{12}$ and $\alpha _{21}$ represent \emph{endogenous} peer effects and $\boldsymbol{\gamma}_{12}$ and $\boldsymbol{\gamma}_{21}$ represent \emph{exogenous} peer effects. From the reduced form, we have
\begin{eqnarray*}
\mathrm{E}(y_{1t}|\mathbf{X}_{1t},\mathbf{X}_{2t}) &=&(1-\alpha _{12}\alpha
_{21})^{-1}\mathbf{X}_{1t}(\boldsymbol{\beta} +\alpha _{12}\boldsymbol{\gamma}_{21})+(1-\alpha _{12}\alpha
_{21})^{-1}\mathbf{X}_{2t}(\boldsymbol{\gamma}_{12}+\alpha _{12}\boldsymbol{\beta} ) \\
\mathrm{E}(y_{2t}|\mathbf{X}_{1t},\mathbf{X}_{2t}) &=&(1-\alpha _{12}\alpha
_{21})^{-1}\mathbf{X}_{2t}(\boldsymbol{\beta} +\alpha _{21}\boldsymbol{\gamma}_{12})+(1-\alpha _{12}\alpha
_{21})^{-1}\mathbf{X}_{1t}(\boldsymbol{\gamma}_{21}+\alpha _{21}\boldsymbol{\beta} ).
\end{eqnarray*}%
Hence, the model is not identified as $\mathrm{E}(y_{it}|\mathbf{X}_{1t},\mathbf{X}_{2t})$ (for $i=1,2$) is perfectly collinear with $\mathbf{X}_{1t}$ and $\mathbf{X}_{2t}$.} A possible restriction to achieve identification is that the researcher knows \emph{a priori} which of the peers' exogenous characteristics directly influence an agent's outcome and which do not.\footnote{We thank an anonymous referee for raising this point.} More specifically, suppose the researcher knows \emph{a priori} that $\mathbf{X}_{it}=[\mathbf{X}^{(1)}_{it},\mathbf{X}^{(2)}_{it}]$ and only $\mathbf{X}^{(1)}_{it}$ can directly affect the outcomes of peers. Then, Equation (\ref{model-mix1}) becomes
\begin{equation*}
y_{it}=\sum_{j \in \mathcal{N}/\{i\}}\alpha _{ij}y_{jt}+\mathbf{X}_{it}\boldsymbol{\beta} +\sum_{j \in \mathcal{N}/\{i\}}\mathbf{X}^{(1)}_{jt}\boldsymbol{\gamma}^{(1)} _{ij} +u_{it},
\end{equation*}%
which can be identified by using $\mathbf{X}^{(2)}_{jt}$ as IVs for $y_{jt}$.
In this case, it may be feasible to test for the presence of endogenous and exogenous peer effects separately. We leave this investigation for future research. $\blacksquare$
\end{remark} 

\begin{remark} [Independent clusters] \label{rmk5}
In general, if the researcher has prior knowledge about which individuals are more likely to be connected in the network, our test can be tailored to assess spillovers specifically among those individuals. This will reduce both the number of restrictions under the null hypothesis and the number of IVs employed to test them. If the prior knowledge aligns well with the true dependence structure, the power of the test is likely to improve.

In some applications, cross-sectional units are located in predetermined groups or clusters (e.g., schools, counties, etc.). When these groups are sufficiently separated (in terms of geographic location or social distance), it is reasonable to assume that spillover effects do not occur across groups.
Suppose the data consist of $R$ disjoint groups with $m_r$ individuals in the $r$th group. Under this structure, the number of restrictions under the null hypothesis, as well as the number of IVs needed for the test, can be substantially reduced to $ \sum^R_{r=1}m_r(m_r-1)$. 
This reduction in the number of restrictions and IVs will improve the power of the test, provided that spillovers indeed occur primarily within groups.

This also relaxes the data requirements implied by Assumption 2. Assumption 2 requires $K<N$, where $N=nT$. In the general case without any prior knowledge about potential peers, $K = n(n-1)$, and thus, the assumption requires $n<T$. In contrast, under the independent cluster setting described above, 
the requirement becomes less restrictive. Let $m_{max}$ denote the size of the largest group, and then $K = \sum^R_{r=1}m_r(m_r-1) < \sum^R_{r=1}m^2_r \leq m_{max}\sum^R_{r=1}m_r = m_{max}\cdot n$. Hence, Assumption 2 is satisfied as long as $m_{max}<T$. In our Monte Carlo simulations, we examine how the presence of independent clusters affects the finite-sample performance of our AR test.\footnote{This assumption, however, should be applied with caution in practice. The gain in power occurs only when the assumption of independent clusters holds. In the extreme case where peer effects exist across clusters but not within them, the test would have no power. In this regard, the AR test that does not rely on such assumptions remains valuable, as it provides a more robust approach when the underlying spillover structure is uncertain.} $\blacksquare$ \end{remark}

\section{AR Test in the Presence of Fixed Effects}\label{ARwf}

To control for unobserved heterogeneity, we introduce individual and time
fixed effects $\xi _{i}$ and $\eta _{t}$ to Equation (\ref{model1}) so that
the error term becomes 
\begin{equation*}
u_{it}=\xi _{i}+\eta _{t}+\epsilon _{it},  \label{fixed_effects}
\end{equation*}%
for $i=1,\cdots ,n$ and $t=1,\cdots ,T$, where $\epsilon _{it}$ are
idiosyncratic random shocks. Then, in matrix form, Equation (\ref{model1_vec}) can
be written as%
\begin{equation}
\mathbf{y}=\mathbf{Y}\boldsymbol{\alpha} +\mathbf{X}\boldsymbol{\beta} +\boldsymbol{\iota} _{T}\otimes \boldsymbol{\xi} +\boldsymbol{\eta} \otimes \boldsymbol{\iota} _{n}+\boldsymbol{\epsilon} ,
\label{model1_fe}
\end{equation}%
where $\boldsymbol{\xi} =(\xi _{1},\cdots ,\xi _{n})^{\prime }$, $\boldsymbol{\eta} =(\eta _{1},\cdots
,\eta _{T})^{\prime }$, and $\boldsymbol{\epsilon} =(\boldsymbol{\epsilon} _{1}^{\prime },\cdots
,\boldsymbol{\epsilon} _{T}^{\prime })^{\prime }$ with $\boldsymbol{\epsilon} _{t}=(\epsilon
_{1t},\cdots ,\epsilon _{nt})^{\prime }$.

To eliminate fixed effects, we apply a two-way within transformation by
premultiplying Equation (\ref{model1_fe}) by $\mathbf{J}=(\mathbf{I}_{T}-T^{-1}\boldsymbol{\iota} _{T}\boldsymbol{\iota}
_{T}^{\prime })\otimes (\mathbf{I}_{n}-n^{-1}\boldsymbol{\iota} _{n}\boldsymbol{\iota} _{n}^{\prime })$. The
transformed model is%
\begin{equation*}
\mathbf{y}^{\ast }=\mathbf{Y}^{\ast }\boldsymbol{\alpha} +\mathbf{X}^{\ast }\boldsymbol{\beta} +\boldsymbol{\epsilon} ^{\ast },
\end{equation*}%
where $\mathbf{y}^{\ast }=\mathbf{Jy}$, $\mathbf{Y}^{\ast }=\mathbf{JY}$, $\mathbf{X}^{\ast }=\mathbf{JX}$, and $\boldsymbol{\epsilon} ^{\ast
}=\mathbf{J}\boldsymbol{\epsilon} $. Moreover, let $\mathbf{Q}^{\ast }$ denote the IV matrix collecting
linearly independent columns in $[\mathbf{X}^{\ast },\mathbf{Z}^{\ast }]$, where $\mathbf{Z}^{\ast }=\mathbf{JZ}$%
, and $\mathbf{P}^{\ast }=\mathbf{Q}^{\ast }(\mathbf{Q}^{\ast \prime }\mathbf{Q}^{\ast })^{-1}\mathbf{Q}^{\ast \prime }$.

The jackknife AR test statistic defined in Equation (\ref{AR1}) has the advantage of being robust to heteroskedasticity of unknown form. However, when individual and time fixed effects exist and a data transformation is used to eliminate these effects, the standard jackknife method no longer re-center the test statistic properly. 
The within transformation introduces both cross-sectional and time-series dependence in the transformed errors and hence 
$\boldsymbol{\epsilon}^{\ast \prime} (\mathbf{P}^{\ast } - \mathbf{D}^{\ast }) \boldsymbol{\epsilon}^{\ast}$, where $\mathbf{D}^{\ast }$ is a diagonal matrix containing the diagonal elements of $\mathbf{P}^{\ast }$, does not have a zero mean. Other transformations such as the Helmert transformation also have the same issue in the presence of heteroskedasticity of unknown form. Hence, instead of using the jackknife method, we maintain the following
assumption regarding the random shocks $\epsilon _{it}$ and re-center the
quadratic form $\boldsymbol{\epsilon} ^{\ast \prime }\mathbf{P}^{\ast }\boldsymbol{\epsilon} ^{\ast }$ by
subtracting out its mean as in \citet{AG2011} and \citet{AS2019}. 

\begin{description}
\item[Assumption 1'] The random shocks $\epsilon _{it}$ are i.i.d. across $i$
and $t$, with $\mathrm{E}(\epsilon _{it}|\mathbf{X}_{it})=0$, $\mathrm{E}(\epsilon
_{it}^{2}|\mathbf{X}_{it})=\sigma ^{2}>0$, and finite eighth conditional moments.
\end{description}

The i.i.d. assumption for $\epsilon_{it}$ may seem restrictive. However, given that it is common practice among empirical researchers to use fixed effects to address potential heterogeneity and correlations in the error term, the i.i.d. assumption -- after accounting for the two-way fixed effects -- is not overly strong.\footnote{In the proof of Proposition \ref{prop2}, we show that the dependence in the transformed errors induced by the within transformation asymptotically vanishes under homoskedasticity. Therefore, the homoskedasticity assumption plays an important role in establishing the asymptotic validity of our test statistic under two-way fixed effects.} As in the case without fixed effects considered in the previous section, we also impose the following assumptions. 

\begin{description}
\item[Assumption 2'] The IV matrix $\mathbf{Q}^{\ast }$ has full column
rank $K^{\ast }$, $K^{\ast }\rightarrow \infty $ as $N\rightarrow
\infty $, and there exists a constant $C_{p}^{\ast }$ such that $
K^{\ast }/N\leq C_{p}^{\ast }<1$.

\item[Assumption 3'] $\mathrm{plim}_{N\rightarrow \infty }\frac{1}{N}\mathbf{X}^{\ast
\prime }\mathbf{X}^{\ast }$ is finite and nonsingular.
\end{description}

Let $N^{\ast }=(n-1)(T-1)$ and $\widehat{\boldsymbol{\epsilon} }^{\ast }=\mathbf{y}^{\ast }-\mathbf{X}^{\ast }\widehat{\boldsymbol{\beta} }$ with $%
\widehat{\boldsymbol{\beta} }=(\mathbf{X}^{\ast \prime }\mathbf{X}^{\ast })^{-1}\mathbf{X}^{\ast \prime }\mathbf{y}^{\ast }$.
The test statistic for $H_{0}:\boldsymbol{\alpha} =\mathbf{0}$ in the presence of fixed effects is%
\begin{equation}
AR_{FE}=\frac{1}{\sqrt{K^{\ast }}\sqrt{\widehat{\Phi }^{\ast }}}\widehat{%
\boldsymbol{\epsilon} }^{\ast \prime }(\mathbf{P}^{\ast }-\frac{K^{\ast }}{N^{\ast }}\mathbf{I}_{N})%
\widehat{\boldsymbol{\epsilon} }^{\ast },  \label{AR2}
\end{equation}%
where $\widehat{\Phi }^{\ast }$ is a consistent estimator of $\Phi ^{\ast
}=(\mu _{4}-3\sigma ^{4})[\mathrm{plim}_{N\rightarrow \infty }\frac{1%
}{K^{\ast }}\sum_{i=1}^N(p_{ii}^{\ast })^{2}-\bar{\lambda}]+2\sigma ^{4}(1-\bar{%
\lambda})$, with $\mu _{4}=\mathrm{E}(\epsilon _{it}^{4}|\mathbf{X}_{it})$, $\bar{%
\lambda}=\lim_{N\rightarrow \infty }K^{\ast }/N$, and $p_{ii}^{\ast }$ being the $i$th diagonal element of $\mathbf{P}^{\ast }$. In Appendix \ref{exk}, we provide a consistent estimator for the excess kurtosis, $\mu _{4}-3\sigma ^{4}$.
It is worth pointing out that, when $\epsilon _{it}$ is mesokurtic (i.e., 
$\mu _{4}-3\sigma ^{4}=0$) or $\mathrm{plim}_{N\rightarrow \infty }%
\frac{1}{K^{\ast }}\sum_{i=1}^N(p_{ii}^{\ast })^{2}=\bar{\lambda}$, we have $%
\Phi ^{\ast }=2\sigma ^{4}(1-\bar{\lambda})$.\footnote{The second case is satisfied when $p_{ii}^{\ast} \to \bar{\lambda}$ for all $i$, which is called an asymptotically balanced design of instruments/regressors in the many-IV literature. See \citet{AY2017} and \citet{AS2019} for related discussions.}

\begin{proposition}
\label{prop2}Suppose Assumptions 1'-3' hold and $\widehat{\Phi }^{\ast }$ is
a consistent estimator of $\Phi ^{\ast }$. Under $H_{0}:\boldsymbol{\alpha} =\mathbf{0}$, the AR
test statistic defined in Equation (\ref{AR2}) is asymptotically standard
normal.
\end{proposition}

As discussed in \citet{AG2011} and \citet{Crudu2021}, the normal approximation does not account for the number of instruments, which can be an issue in finite samples, particularly when the number of instruments is relatively small. Therefore, we consider the following chi-square approximation, valid when $K^{\ast}$ is either fixed or goes to infinity, and use it for our Monte Carlo simulations and empirical applications.
Let $q_{f}(\tau)$ denote the $\tau$th quantile of the chi-square distribution with $f$ degrees of freedom.
\begin{corollary}
\label{corol} Suppose (i) $K^{\ast}$ goes to infinity and the assumptions of Proposition \ref{prop2} hold, or (ii) $K^{\ast}$ is fixed, $\mathrm{plim}_{N\rightarrow \infty }\frac{1}{N}\mathbf{X}^{\ast \prime}\mathbf{Q}^{\ast}$ is finite with rank $L$, 
$\mathrm{plim}_{N\rightarrow \infty }\frac{1}{N}\mathbf{Q}^{\ast \prime }\mathbf{Q}^{\ast }$ is finite and nonsingular, 
and $\mathbf{Q}^{\ast \prime }\boldsymbol{\epsilon}/\sqrt{N} \overset{d}{\rightarrow} N(\mathbf{0},\sigma^2\mathrm{plim}_{N\rightarrow \infty }\frac{1}{N}\mathbf{Q}^{\ast \prime }\mathbf{Q}^{\ast })$. Then, under $H_{0}:\boldsymbol{\alpha} =\mathbf{0}$, 
\begin{equation*}
\textnormal{Pr} \left( \sqrt{2K^{\ast }}AR_{FE} + K^{\ast } \geq q_{K^{\ast } - L} \left( 1 - \tau \right) \right) \rightarrow \tau,
\end{equation*} where $L$ is the number of columns in $\mathbf{X}$.
\end{corollary}

\begin{remark} \label{rmk6} When network links are unobserved, as an ad-hoc solution, researchers often assume that each individual is \textit{equally} influenced by \textit{all the other} individuals in the network. This is known as the linear-in-means model:
\begin{equation*}
y_{it}=\rho\frac{1}{n-1}\sum_{j\neq i}y_{jt}+\mathbf{X}_{it}\boldsymbol{\beta} +u_{it}.
\end{equation*}%
However, in the presence of fixed effects $u_{it}=\xi _{i}+\eta _{t}+\epsilon _{it}$, the peer effect coefficient $\rho$ is not identifiable after the within transformation \citep[e.g.,][]{Lee07group,BDF09}. To be more specific, the linear-in-means model can be written in matrix form:
\begin{equation*}
\mathbf{y}=\rho (\mathbf{I} _{T}\otimes\mathbf{W}_{\text{LIM}})\mathbf{y}+\mathbf{X}\boldsymbol{\beta} +\boldsymbol{\iota} _{T}\otimes \boldsymbol{\xi} +\boldsymbol{\eta} \otimes \boldsymbol{\iota} _{n}+\boldsymbol{\epsilon} ,
\end{equation*}%
where $\mathbf{W}_{\text{LIM}}=(\boldsymbol{\iota}_{n}\boldsymbol{\iota} _{n}^{\prime }-\mathbf{I}_{n})/(n-1)$ is a zero-diagonal matrix with all off-diagonal elements being $1/(n-1)$.
As $\mathbf{J}[\rho(\mathbf{I} _{T}\otimes\mathbf{W}_{\text{LIM}})]= \rho^{\ast }\mathbf{J}$, where $\rho^{\ast } = -\rho/(n-1)$,
premultiplying the model by $\mathbf{J}$ gives%
\begin{equation*}
\mathbf{y}^{\ast }=\rho^{\ast } \mathbf{y}^{\ast }+\mathbf{X}^{\ast }\boldsymbol{\beta} +\boldsymbol{\epsilon}^{\ast } ,
\end{equation*}%
with the reduced form 
\begin{equation*}
\mathbf{y}^{\ast }=(1-\rho^{\ast } )^{-1}(\mathbf{X}^{\ast }\boldsymbol{\beta} +\boldsymbol{\epsilon} ^{\ast }).
\end{equation*}%
Hence, in the presence of fixed effects, $\rho$ cannot be separately identified from $\boldsymbol{\beta}$ based on the conditional mean of $\mathbf{y}^{\ast }$ in the linear-in-means model.\footnote{An alternative interpretation of this non-identification result is as follows. To estimate the linear-in-means model by the two-stage least squares, possible IVs for the endogenous regressor $ (\mathbf{I} _{T}\otimes\mathbf{W}_{\text{LIM}})\mathbf{y}$ are $ (\mathbf{I} _{T}\otimes\mathbf{W}_{\text{LIM}})\mathbf{X}$, $ (\mathbf{I} _{T}\otimes\mathbf{W}^{2}_{\text{LIM}})\mathbf{X}$, etc. However, after the within transformation, all these IVs are linearly dependent on $\mathbf{X}^{\ast}$ and hence the model cannot be identified.}
In contrast, the proposed AR test 
can detect peer effects as long as (i) the true network is not complete or (ii) the true network is complete but the peer effects are heterogeneous.\footnote{A network is complete if all individuals in the network are linked with each other.} The  Monte Carlo simulations reported in Section \ref{simul} demonstrate this important advantage of our test. $\blacksquare$ 
\end{remark}

\section{Monte Carlo Simulations}\label{simul}

In this section, we examine the empirical size and power of the AR test proposed in Equation (\ref{AR2}) (hereafter, denoted as $\texttt{T}_{\texttt{JL}}$) using simulations.\footnote{To save space, we do not report the results of the simulations without fixed effects. The results are available upon request.} The estimator proposed in Appendix \ref{exk} is used to calculate the excess kurtosis in $\texttt{T}_{\texttt{JL}}$. All test statistics considered in this section use the same set of two-way within transformed residuals described in Section \ref{ARwf}. Appendix \ref{addsimul} includes additional simulation results that are not reported in this section. The number of repetitions for each simulation specification is 5,000.

\subsection{Size and Power}
In the simulations, the data are generated from
\begin{equation}
y_{it}=\sum_{j \in \mathcal{N}/\{i\}}\alpha _{ij}y_{jt}+\beta x_{it} +u_{it},
\label{DGP}
\end{equation}
where $x_{it} \sim \text{i.i.d.N}(0,1)$ and $u_{it} = \xi_i + \eta_t + \epsilon_{it}$ with $\xi_i \sim \text{i.i.d.U}(-1,1)$ and $\eta_t \sim \text{i.i.d.U}(-1,1)$. The random errors $\epsilon_{it}$ are generated from either normal:  $\epsilon_{it} \sim \text{i.i.d.N}(0,1)$ (\texttt{DGP1}) or  log-normal: $\epsilon_{it} \sim \text{i.i.d.}$ $[\exp(\zeta_{it})-\exp(0.5)]/[\exp(2)-\exp(1)]^{0.5}$ with $\zeta_{it} \sim \text{i.i.d.N}(0,1)$ (\texttt{DGP2}). 

The true network is generated as a random graph \citep{er1959}. Let $\texttt{ND}$ denote the proportion of dyads in the network with non-zero peer effect coefficients. Thus, $\texttt{ND}$ represents the density of the underlying network. We randomly select $\texttt{ND} \times 100\%$ of the dyads and set the corresponding peer effect coefficients $\alpha_{ij}=\rho$. \\

\noindent \textbf{SIZE} \, To demonstrate the robustness of the proposed AR test, $\texttt{T}_{\texttt{JL}}$, to many IVs, we compare $\texttt{T}_{\texttt{JL}}$ with two existing tests in the many-IV literature: \citet{Donald2003}'s J test ($\texttt{T}_{\texttt{DIN}}$), which is essentially $\texttt{T}_{\texttt{JL}}$ with the variance term $\Phi^{\ast}$ being $2\sigma^4$ due to the assumption of moderately many IVs such that $K^2/N \to 0$ \citep{AG2011}; and \citet{AG2011}'s J test ($\texttt{T}_{\texttt{AG}}$), where $\Phi^{\ast} = 2\sigma^4(1 - \bar{\lambda})$ with $\bar{\lambda} = \lim_{N \to \infty} K^{\ast }/N$ due to the balanced covariate design assumption that $p_{ii}^{\ast} \to \bar{\lambda}$ for all $i$. All three tests use the same set of IVs described in Sections \ref{ARwof}. Chi-square approximation, described in Section \ref{ARwf}, is used to calculate the critical values.

\begin{table}[h!]
\centering
  \begin{threeparttable}
 \caption{Empirical Size} \label{tb:1}
\begin{tabular}{rrc|p{1.2cm}p{1.2cm}p{1.2cm}p{1.2cm}p{1.2cm}p{1.2cm}}
\hline \hline
   &    &           & \multicolumn{2}{c}{$\texttt{T}_{\texttt{DIN}}$} & \multicolumn{2}{c}{$\texttt{T}_{\texttt{AG}}$} & \multicolumn{2}{c}{$\texttt{T}_{\texttt{JL}}$} \\
$n$  & $T$  & \texttt{DGP} &  5\%        & 1\%      & 5\%  & 1\%  & 5\%   & 1\%  \\
\hline
5  & 50  & 1 & 0.039 & 0.006 & 0.043 & 0.007 & 0.043 & 0.007 \\
10 & 50  & 1 & 0.028 & 0.005 & 0.044 & 0.008 & 0.044 & 0.008 \\
20 & 50  & 1 & 0.017 & 0.001 & 0.046 & 0.007 & 0.046 & 0.007 \\
30 & 50  & 1 & 0.003 & 0.000 & 0.037 & 0.005 & 0.037 & 0.005 \\
40 & 50  & 1 & 0.000 & 0.000 & 0.032 & 0.003 & 0.032 & 0.003 \\
5  & 100 & 1 & 0.044 & 0.008 & 0.045 & 0.008 & 0.046 & 0.008 \\
10 & 100 & 1 & 0.043 & 0.006 & 0.050 & 0.008 & 0.050 & 0.008 \\
20 & 100 & 1 & 0.038 & 0.005 & 0.052 & 0.009 & 0.052 & 0.009 \\
30 & 100 & 1 & 0.026 & 0.002 & 0.046 & 0.010 & 0.047 & 0.010 \\
40 & 100 & 1 & 0.018 & 0.002 & 0.049 & 0.009 & 0.049 & 0.009 \\ \hline
5  & 50  & 2 & 0.060 & 0.014 & 0.066 & 0.016 & 0.045 & 0.007 \\
10 & 50  & 2 & 0.064 & 0.018 & 0.082 & 0.028 & 0.047 & 0.009 \\
20 & 50  & 2 & 0.047 & 0.010 & 0.088 & 0.031 & 0.041 & 0.008 \\
30 & 50  & 2 & 0.022 & 0.004 & 0.089 & 0.029 & 0.040 & 0.005 \\
40 & 50  & 2 & 0.004 & 0.000 & 0.088 & 0.028 & 0.031 & 0.002 \\
5  & 100 & 2 & 0.064 & 0.016 & 0.067 & 0.018 & 0.053 & 0.011 \\
10 & 100 & 2 & 0.061 & 0.016 & 0.068 & 0.019 & 0.045 & 0.008 \\
20 & 100 & 2 & 0.069 & 0.019 & 0.090 & 0.028 & 0.054 & 0.011 \\
30 & 100 & 2 & 0.051 & 0.013 & 0.083 & 0.028 & 0.045 & 0.008 \\
40 & 100 & 2 & 0.042 & 0.010 & 0.086 & 0.029 & 0.044 & 0.009 \\
\hline \hline
\end{tabular}
\begin{tablenotes}
 \small
 \item Note: $\texttt{T}_{\texttt{JL}}$ is the proposed AR test. $\beta = 1$, \texttt{DGP1} - normal errors, and \texttt{DGP2} - log-normal errors. All tests use the same set of IVs and two-way within-transformed residuals described in Sections \ref{ARwof} and \ref{ARwf}. The Chi-square distribution is used to calculate the critical values.
   \end{tablenotes}
    \end{threeparttable}
\end{table}

Table \ref{tb:1} reports the empirical size of the three tests for nominal 5\% and 1\% significance levels. When $n$ is large relative to $T$, which corresponds to the case where the number of IVs increases as fast as the sample size (i.e., $\bar{\lambda} \approx 1$),\footnote{In the simulations, $K^{\ast} = n(n-1) + 1$ and thus, $\bar{\lambda} \equiv K^{\ast}/N \approx 1$ when $n \approx T$, where $N = nT$.} $\texttt{T}_{\texttt{DIN}}$ exhibits significant under-rejections, while the rejection rates of the other tests are much closer to the nominal rates. This aligns with the theoretical prediction in \citet[][Theorem 1]{AG2011} that $\texttt{T}_{\texttt{DIN}}$ under-rejects the null when $\bar{\lambda} > 0$ as the sample size increases. When $n$ is close to $T$, $\texttt{T}_{\texttt{JL}}$ slightly under-rejects the null, but the under-rejection vanishes as $T$ increases.

When the random error $\epsilon_{it}$ is normal, $\texttt{T}_{\texttt{AG}}$ and $\texttt{T}_{\texttt{JL}}$ show almost the same rejection rate, which is because the excess kurtosis in this case is zero and thus the variance term $\Phi^{\ast}$ in $\texttt{T}_{\texttt{JL}}$ reduces to the $\texttt{T}_{\texttt{AG}}$'s variance, $2\sigma^4(1 - \bar{\lambda})$. When the error is log-normal, however, the additional variance components in $\Phi^{\ast}$, that are associated with the excess kurtosis and the diagonal elements of the projection matrix, do not vanish, creating additional sampling variability that $\texttt{T}_{\texttt{AG}}$ does not account for. Thus, $\texttt{T}_{\texttt{AG}}$ exhibits significant over-rejections in this case. By contrast, the rejection rates of $\texttt{T}_{\texttt{JL}}$ are much closer to the nominal rates. These results also indicate that the estimator for the excess kurtosis proposed in Appendix \ref{exk} performs well. \\

\begin{figure}[H]
     \centering
            \caption{Power Curves of the AR Test with Varying $\rho$ and $\beta$}
        \label{fig:power1}
     \begin{subfigure}[b]{0.495\textwidth}
         \centering
         \includegraphics[width=\textwidth]{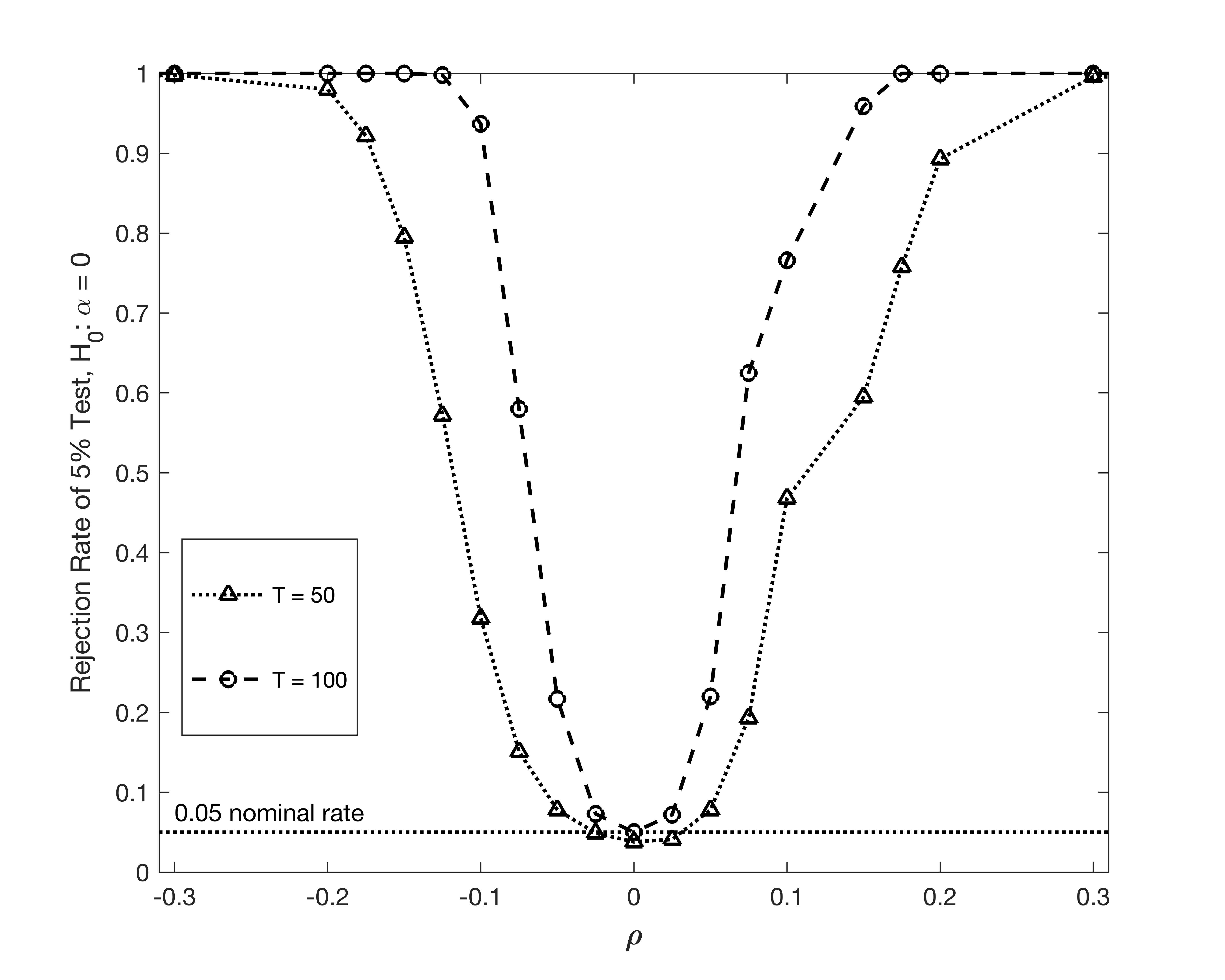}
         \caption{Varying $\rho$ with $\beta = 1$, Normal Errors}
     \end{subfigure}
     \begin{subfigure}[b]{0.495\textwidth}
         \centering
         \includegraphics[width=\textwidth]{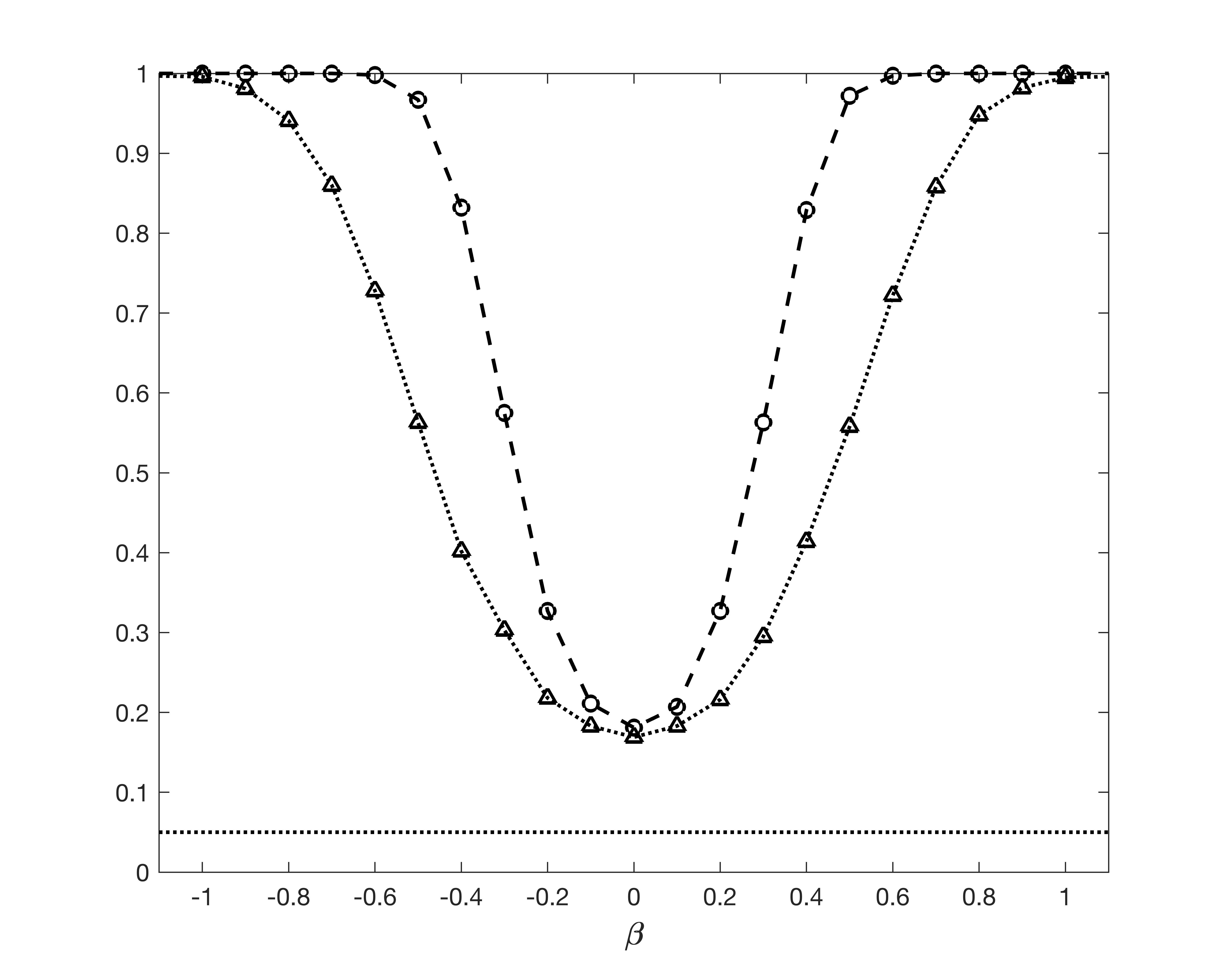}
         \caption{Varying $\beta$ with $\rho$ = 0.3, Normal Errors}
     \end{subfigure}
          \begin{subfigure}[b]{0.495\textwidth}
         \centering
         \includegraphics[width=\textwidth]{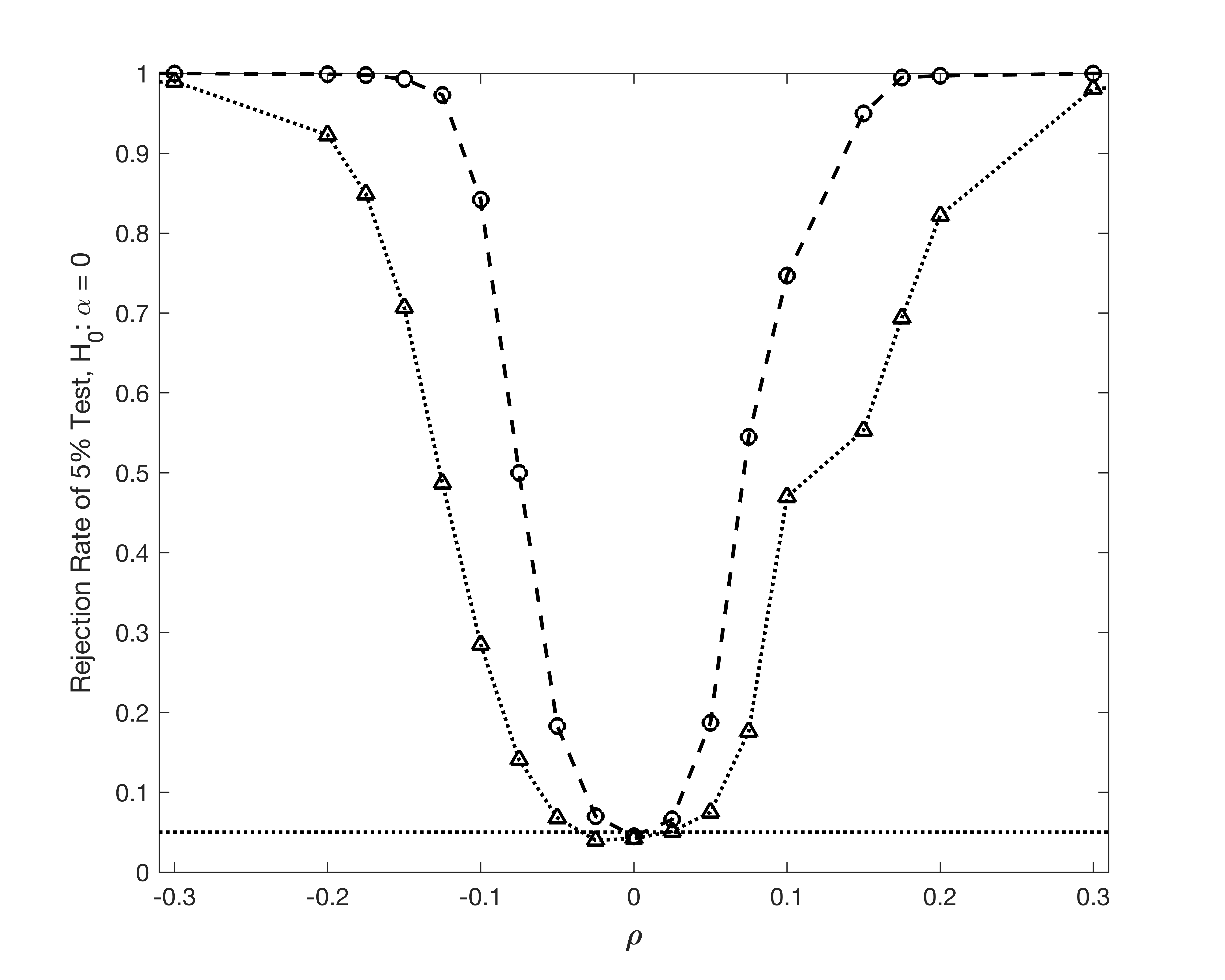}
         \caption{Varying $\rho$ with $\beta = 1$, Log-Normal Errors}
     \end{subfigure}
     \begin{subfigure}[b]{0.495\textwidth}
         \centering
         \includegraphics[width=\textwidth]{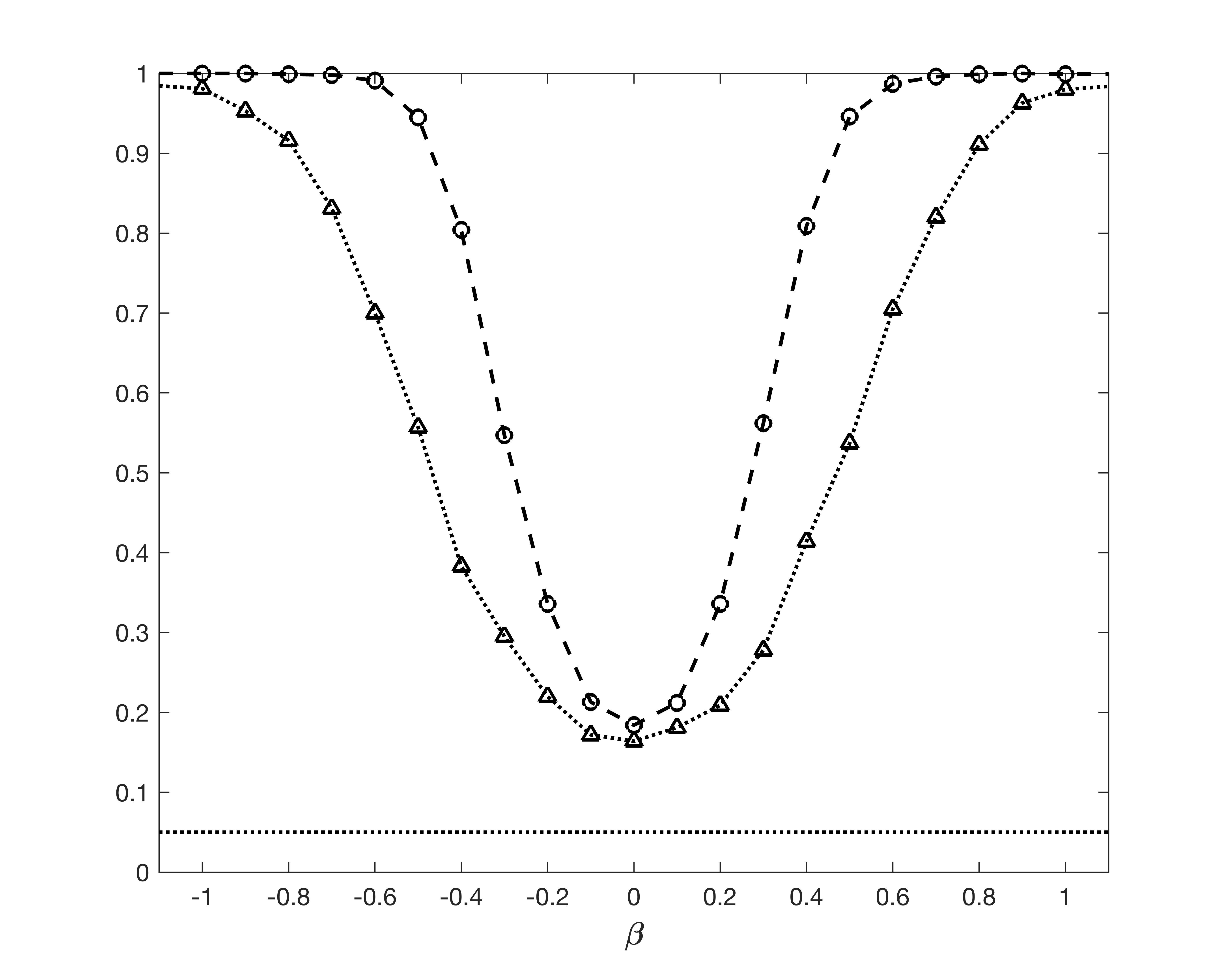}
         \caption{Varying $\beta$ with $\rho = 0.3$, Log-Normal Errors}
     \end{subfigure}
     \caption*{\small Note: $n=30$, $\texttt{ND} = 0.3$, and $5\%$ significance level test. The Chi-square distribution is used to calculate the critical values.}
\end{figure}

\noindent \textbf{POWER} \, 
Figure \ref{fig:power1} presents the power curves of the proposed AR test with varying $\rho$ and $\beta$, when $\texttt{ND} = 0.3$, $n=30$, and $T \in \{50,100\}$. 

The rejection rates of the test quickly increase as the absolute values of $\rho$ and $\beta$ increase, confirming the analytic prediction in Remark \ref{power}. Overall, it appears that the power is close to unity when $\rho = 0.3$, $\beta = 1$, 
and $T=50$.\footnote{Appendix \ref{addsimul} compares the power of $\texttt{T}_{\texttt{AG}}$ and $\texttt{T}_{\texttt{JL}}$, where they exhibit almost the same rejection rates in many settings, indicating that there is little or no power loss when using $\texttt{T}_{\texttt{JL}}$, instead of $\texttt{T}_{\texttt{AG}}$, for testing the presence of peer effects without the balanced covariate design assumption.}

The next simulations analyze how network density \texttt{ND} affects the power of the test. The second column of Table \ref{tb:power_ND} reports the rejection rates of the AR test under different levels of network density \texttt{ND}. Overall, the power of the AR test increases as \texttt{ND} increases. However, when the network is sparse (e.g., \texttt{ND} $\leq 0.02$),\footnote{In these simulations, $n = 30$, so the number of possible links is $30 \times 29 = 870$. When \texttt{ND} = 0.02, approximately 17 nodes out of 870 are nonzero.} the power of the AR test is low. 

\begin{table}[H] \singlespacing 
\centering
\begin{threeparttable} 
\caption{Network Density and the Power of AR test}
\label{tb:power_ND}
\begin{tabular}{c|c|cc} \hline \hline
     & $\texttt{T}_{\texttt{JL}}$ & \multicolumn{2}{c}{$\widehat{\beta}$}                            \\
\texttt{ND}   & 5\%   & Bias/true value    & 95\% CI  \\ \hline
0.01 & 0.113 & -0.0027   & 0.949        \\
0.02 & 0.256 & -0.0067   & 0.941        \\
0.03 & 0.403 & -0.0096   & 0.936        \\
0.04 & 0.542 & -0.0115   & 0.928        \\
0.05 & 0.643 & -0.0152   & 0.910        \\
0.06 & 0.692 & -0.0165   & 0.914        \\
0.07 & 0.703 & -0.0194   & 0.906        \\
0.08 & 0.680 & -0.0226   & 0.897        \\
0.09 & 0.634 & -0.0226   & 0.904        \\
0.10 & 0.584 & -0.0270   & 0.924        \\
0.20 & 0.956 & -0.0482   & 0.712        \\
0.30 & 0.997 & -0.0685   & 0.541       
  \\ \hline \hline
\end{tabular}
\begin{tablenotes}
 \small
 \item Note: $(n,T) = (30,50),$ $\beta = 1$, $\rho = 0.3$ and $5\%$ significance level test. ``95\% CI" indicates the coverage rate of 95\% confidence intervals of $\widehat{\beta}$.
  \end{tablenotes}
  \end{threeparttable}
\end{table}

Note that the peer effect coefficients $\alpha_{ij}$ are fixed to $\rho = 0.3$ in these simulations, so the overall level of peer interactions in the model is quite low when \texttt{ND} $\leq 0.02$.
Since the proposed testing approach tests for all possible network connections without targeting any specific network structure, the low rejection rate in this scenario is expected (see Remark \ref{power} for more discussion). In this case, shrinkage methods that exploit the sparsity structure of the network may be preferable.  Given that shrinkage methods may not perform well with dense networks, the two methods may complement each other in practice.\footnote{It is worth pointing out that the identification of both network links and peer effect parameters in \citet{DRS24} is considerably more challenging than testing for the existence of peer effects in our approach. Consequently, the identification strategy in \citet{DRS24} requires stronger regularity conditions. In particular, it requires that network effects do not cancel out such that $\beta_0\rho_0 + \gamma_0 \ne 0$ (Assumption A3), which excludes the null of no peer effects, i.e., $\rho_0=0$ and $\gamma_0=0$. As a result, their identification strategy and the adaptive elastic net GMM estimator developed under this assumption may not be suitable for testing the null of no peer effects directly.}

As discussed in Introduction, we envision the proposed AR test to be useful in scenarios where the primary parameter of interest is $\boldsymbol{\beta}$ and the researcher estimates $\boldsymbol{\beta}$ under the assumption of no network effects. The researcher can use our test to provide supportive evidence for the validity of the estimates.
We would like to argue that the low rejection rates of the proposed AR test with a sparse network may not be a significant issue in this context. 
The third and fourth columns of Table \ref{tb:power_ND} report the bias and the coverage rate of 95 \% confidence intervals (CI) of $\widehat{\beta }=(\mathbf{x}^{\ast \prime }\mathbf{x}^{\ast })^{-1}\mathbf{x}^{\ast \prime }\mathbf{y}^{\ast }$. 
As $\widehat{\beta}$ is a regression estimate of $\beta$ without accounting for peer effects, the bias and the distortion of the coverage rate of $\widehat{\beta}$ can represent the cost of type \rom{2} error. When $\texttt{ND}$ is less than 4\%, the rejection rate is less than 50\%, but the under-coverage of the 95\% CI of $\widehat{\beta}$ is still less than 2\%. 
If this level of bias and under-coverage is acceptable in practice, our test can still appeal to empirical researchers who prefer methods with less restrictive assumptions and simpler computations.

Next, we explore the idea of independent clusters discussed in Remark \ref{rmk5}. The simulations are based on the design used in Table \ref{tb:power_ND}, with the modification that the $n$ individuals are divided into $R$ equal-sized groups. We then randomly select \texttt{ND} × 100\% of the dyads within each group and set the corresponding $\alpha_{ij}=0.3$.
As a result, the simulated network structure conforms to the assumption of independent clusters. We consider two tests, $\texttt{T}_{\texttt{JL}}$ and $\widetilde{\texttt{T}}_{\texttt{JL}}$. The former does not impose the assumption of independent clusters, whereas the latter is constructed under that assumption.
In these simulations, we set $n = 30$. Consequently, the number of IVs in $\texttt{T}_{\texttt{JL}}$ is fixed at $K=n(n-1) = 870$ and that in $\widetilde{\texttt{T}}_{\texttt{JL}}$ varies with $R$ such that $K=n(n/R-1)$. For example, when $R = 2$, the number of IVs in $\widetilde{\texttt{T}}_{\texttt{JL}}$ is 420 and this number drops significantly to 60 when $R = 10$. The simulation results are presented in Table \ref{tb:cluster}.

\begin{table}[H] \singlespacing 
\centering
\begin{threeparttable} 
\caption{Independent Clusters and the Power}
\label{tb:cluster}
\begin{tabular}{c@{\hskip 0.2in}|c@{\hskip 0.2in}c@{\hskip 0.1in}|c@{\hskip 0.2in}c@{\hskip 0.1in}|c@{\hskip 0.2in}c@{\hskip 0.1in}|c@{\hskip 0.2in}c@{\hskip 0.1in}} \hline \hline
& \multicolumn{2}{c|}{\texttt{ND} = 0.00} & \multicolumn{2}{c|}{\texttt{ND} = 0.01} & \multicolumn{2}{c|}{\texttt{ND} = 0.02} & \multicolumn{2}{c}{\texttt{ND} = 0.03} \\ 
 $R$ & $\texttt{T}_{\texttt{JL}}$ & $\widetilde{\texttt{T}}_{\texttt{JL}}$ & $\texttt{T}_{\texttt{JL}}$ & $\widetilde{\texttt{T}}_{\texttt{JL}}$ & $\texttt{T}_{\texttt{JL}}$ & $\widetilde{\texttt{T}}_{\texttt{JL}}$ & $\texttt{T}_{\texttt{JL}}$ & $\widetilde{\texttt{T}}_{\texttt{JL}}$ \\ \hline
2               & 0.036  & 0.049 & 0.131     & 0.328 & 0.247   & 0.677 & 0.490   & 0.933 \\
3               & 0.037  & 0.042 & 0.116     & 0.413 & 0.261   & 0.842 & 0.422   & 0.977 \\
6               & 0.037  & 0.047 & 0.150     & 0.830 & 0.239   & 0.969 & 0.587   & 1.000 \\
10              & 0.039  & 0.051 & 0.127     & 0.854 & 0.262   & 0.997 & 0.498   & 1.000    \\ \hline \hline
\end{tabular}
\begin{tablenotes}
 \small
 \item Note: $(n,T) = (30,50),$ $\beta = 1$, $\rho = 0.3$, normal errors, and $5\%$ significance level test. $R$: the number of independent clusters.
  \end{tablenotes}
  \end{threeparttable}
\end{table}

Improvements in both size and power are observed for $\widetilde{\texttt{T}}_{\texttt{JL}}$. The improvement in size may be attributed to the chi-square approximation used to calculate the critical values. As shown in the corollary to Proposition 2, the approximation is valid when the number of IVs is either fixed or grows as fast as the sample size, but may yield better finite sample results when the number of IVs is relatively small.

\subsection{Comparison against Tests with Misspecified Networks}\label{mis}

An important merit of the proposed AR test is that it is not contingent on any specific form of network structure so it can be more robust compared to existing tests based on potentially misspecified networks. To show this merit, we compare the proposed AR test ($\texttt{T}_{\texttt{JL}}$) for $H_0:\alpha_{ij}=0$ for all $(i,j)$ in the model defined in Equation (\ref{DGP}) against a t-test ($\texttt{t-test}_{\texttt{TSLS}}$) for $H_0:\rho=0$ based on the two-stage least squares (TSLS) estimation of the model 
\begin{equation}
\mathbf{y}=\rho (\mathbf{I} _{T}\otimes\mathbf{W})\mathbf{y}+\mathbf{X}\boldsymbol{\beta} +\boldsymbol{\iota} _{T}\otimes \boldsymbol{\xi} +\boldsymbol{\eta} \otimes \boldsymbol{\iota} _{n}+\boldsymbol{\epsilon}, \label{model_sarfe}
\end{equation}%
where the adjacency matrix $\mathbf{W}$ is potentially misspecified. To carry out the TSLS estimation, we apply the two-way within transformation to obtain
\begin{equation*}
\mathbf{Jy}=\rho \mathbf{J}(\mathbf{I} _{T}\otimes\mathbf{W})\mathbf{y}+\mathbf{JX}\boldsymbol{\beta}+ \mathbf{J}\boldsymbol{\epsilon},
\end{equation*}
and use $\mathbf{J}(\mathbf{I} _{T}\otimes\mathbf{W})\mathbf{X}$ as the IV for $\mathbf{J}(\mathbf{I} _{T}\otimes\mathbf{W})\mathbf{y}$.

We consider three cases of misspecification in the network structure that are commonly encountered in empirical research: size, location and direction.
The first case misspecifies the sizes of the network links. 
In this case, the true network is generated as a random graph, where we randomly select $\texttt{ND} \times 100\%$ of the dyads and set the corresponding coefficients $\alpha_{ij} \sim \text{U}(0,1)$. To obtain the misspecified adjacency matrix $\mathbf{W}$ in Equation (\ref{model_sarfe}), we define an indicator matrix $\mathbf{W}^{\ast}=[w_{ij}^{\ast}]$, where $w_{ij}^{\ast} = \mathds{1}\{ \alpha_{ij} > 0 \}$ for $i \ne j$, and then row-normalize $\mathbf{W}^{\ast}$ to get $\mathbf{W}$. 
This corresponds to the situation where a researcher knows the locations of the non-zero network links (i.e., who is connected with whom in the network) but does not know the size (or strength) of the links, so assumes each node is equally influenced by all its connections as an ad hoc solution.\footnote{An alternative interpretation is that the peer effects are heterogeneous in the data-generating process but the researcher (mistakenly) assumes that the peer effect is homogeneous.}
We set $(n,T) = (30,50)$ and $\beta = 1$, and experiment with different values of $\texttt{ND}$. Note that, as $\texttt{ND}$ determines the number of non-zero links in the network, it also determines the number of misspecified entries in $\mathbf{W}$. Hence, $\texttt{ND}$ captures the degree of misspecification.

\begin{table}[H] \singlespacing
\centering
\begin{threeparttable}
\caption{Misspecification in the Sizes of Network Links} \label{tb:2}
\begin{tabular}{c|cp{2cm}p{2cm}|p{2cm}p{2cm}} \hline \hline
     & & \multicolumn{2}{c|}{Normal Errors} & \multicolumn{2}{c}{Log-Normal Errors} \\
$\texttt{ND}$   & & $\texttt{T}_{\texttt{JL}}$    & $\texttt{t-t}_{\texttt{TSLS}}$  & $\texttt{T}_{\texttt{JL}}$    & $\texttt{t-t}_{\texttt{TSLS}}$  \\ \hline
0.01 & & 0.504 & 1.000 & 0.435 & 1.000 \\
0.10  && 0.838 & 0.951 & 0.847 & 0.948 \\
0.30  && 0.991 & 0.641 & 0.982 & 0.655 \\
0.50  && 0.993 & 0.487 & 0.987 & 0.491 \\
0.70  && 0.990 & 0.430 & 0.983 & 0.433 \\
0.90  && 0.993 & 0.399 & 0.983 & 0.394 \\
0.93 && 0.992 & 0.372 & 0.987 & 0.382 \\
0.96 && 0.991 & 0.287 & 0.985 & 0.307 \\
0.99 & &0.994 & 0.087 & 0.988 & 0.091 \\
1.00    && 0.993 & N.A.    & 0.986 & N.A. \\ \hline \hline
\end{tabular}
\begin{tablenotes}
 \small
 \item Note: $(n,T) = (30,50),$ $\beta = 1,$ and $5\%$ significance test. ``N.A." indicates ``not applicable." 
  \end{tablenotes}
  \end{threeparttable}
\end{table}

Table \ref{tb:2} reports the rejection rates of $\texttt{T}_{\texttt{JL}}$ and $\texttt{t-test}_{\texttt{TSLS}}$ under varying $\texttt{ND}$. The power of $\texttt{t-test}_{\texttt{TSLS}}$ is higher than the power of $\texttt{T}_{\texttt{JL}}$ when $\texttt{ND}$ is small, but as $\texttt{ND}$ increases, the power of $\texttt{T}_{\texttt{JL}}$ improves, whereas the power of $\texttt{t-test}_{\texttt{TSLS}}$ deteriorates. 
This is because, as $\texttt{ND}$ increases, the overall peer effect in the model increases but the number of misspecified links in $\mathbf{W}$ also increases -- the former increases the power of $\texttt{T}_{\texttt{JL}}$ while the latter decreases the power of $\texttt{t-test}_{\texttt{TSLS}}$.
When $\texttt{ND}$ is close to one, $\texttt{t-test}_{\texttt{TSLS}}$ has little power and eventually is not applicable. This is because, as $\texttt{ND}$ gets close to one, the misspecified network $\mathbf{W}$ converges to a complete and homogeneous network, and the model defined in Equation (\ref{model_sarfe}) becomes the linear-in-means model. 
As discussed in Remark \ref{rmk6}, in the presence of fixed effects, the peer effect coefficient $\rho$ cannot be identified in the linear-in-means model. In contrast, the proposed AR test performs well even if the true network is complete as long as the true $\alpha_{ij}$'s are heterogeneous.

The second case pertains to misspecification in the locations of non-zero network links. In this experiment, the true network is a circular network where all individuals are equally spaced around a circle and are friends only with their two nearest neighbors. 
We assume individuals are equally influenced by their friends.
Specifically, in Equation (\ref{DGP}), we set $\alpha_{ij} = 0.3$ if $j$ is $i$'s nearest neighbor (i.e., right next to $i$) and $\alpha_{ij} = 0$ otherwise.
To obtain the misspecified adjacency matrix $\mathbf{W}=[w_{ij}]$ in Equation (\ref{model_sarfe}), we assume the researcher mistakenly takes the $\texttt{m}^{\text{th}}$ nearest neighbors as one's friends. Specifically, $w_{ij} = 0.5$ if $j$ is $i$'s $\texttt{m}^{\text{th}}$ nearest neighbor and $w_{ij} = 0$ otherwise.\footnote{Note that each individual always has two friends, so $w_{ij} = 0.5$ is obtained when the adjacency matrix is row-normalized.}
Therefore, as $\texttt{m}$ increases, the degree of misspecification increases. We set $(n,T) = (30,50)$ and $\beta=1$, and experiment with different values of $\texttt{m}$.

\begin{figure}[H]
     \centering
            \caption{Misspecification in the Locations of Non-Zero Network Links}
        \label{fig:power3}
     \begin{subfigure}[b]{0.495\textwidth}
         \centering
         \includegraphics[width=\textwidth]{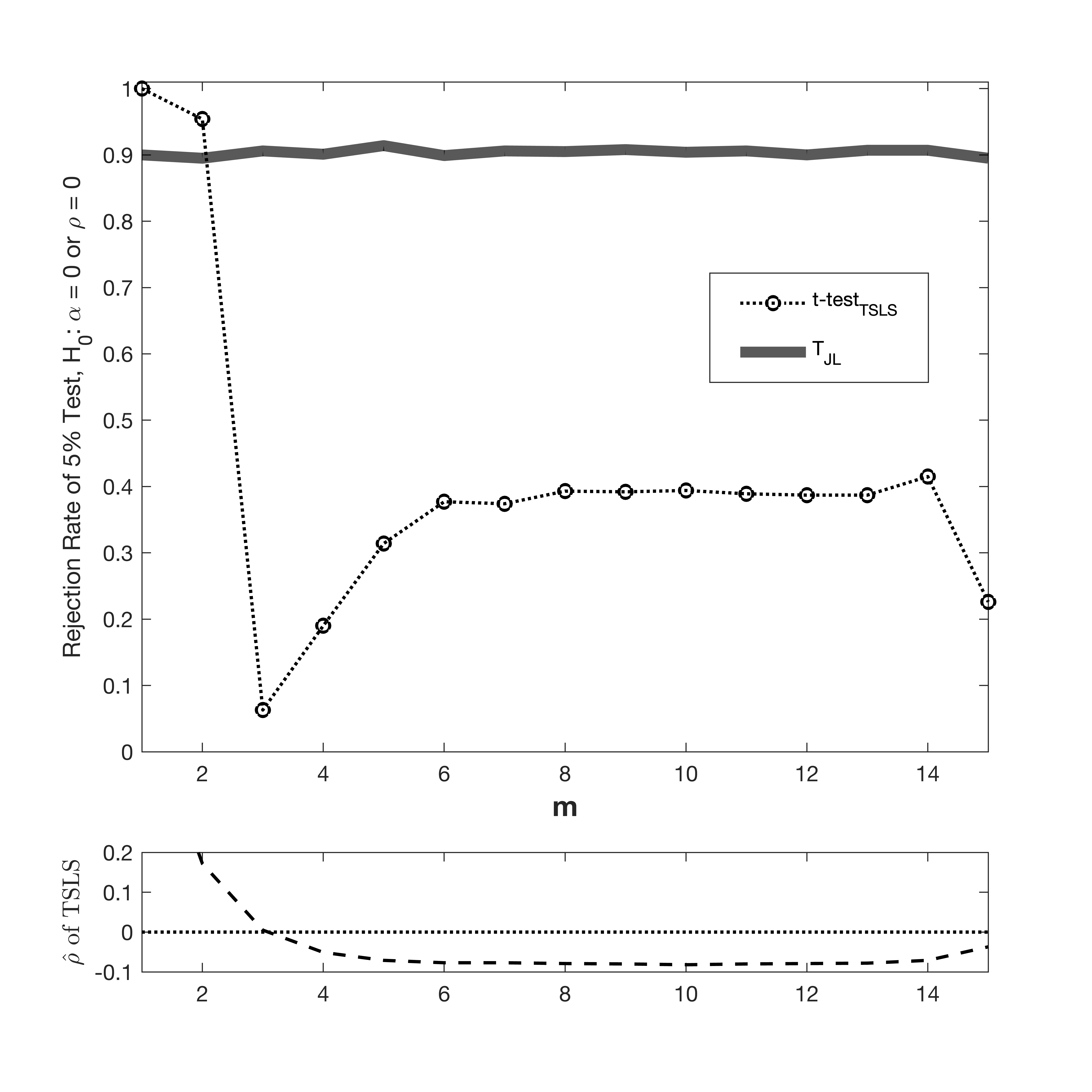}
         \caption{Normal Errors}
     \end{subfigure}
     \begin{subfigure}[b]{0.495\textwidth}
         \centering
         \includegraphics[width=\textwidth]{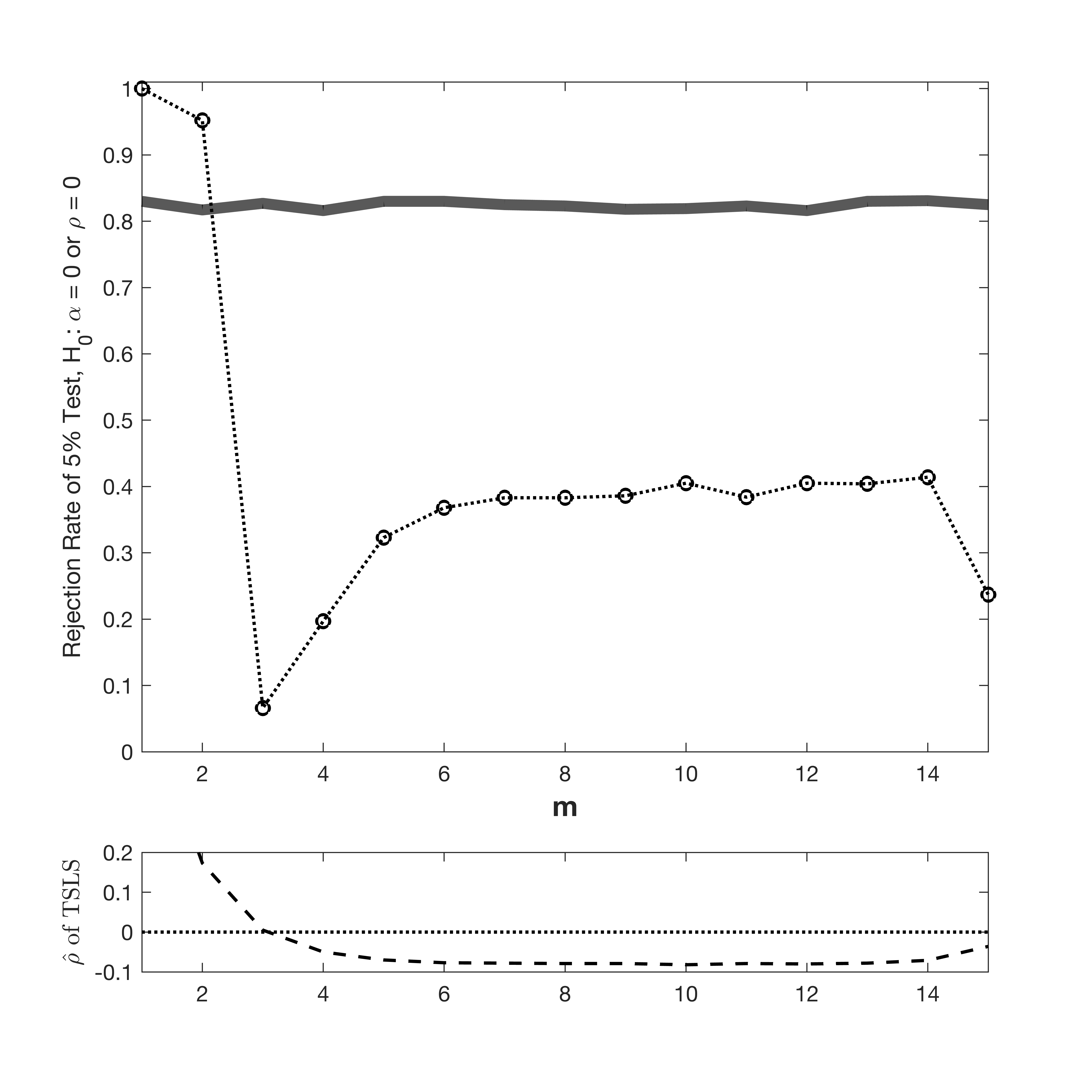}
         \caption{Log-Normal Errors}
     \end{subfigure}
     \caption*{\small Note: $(n,T) = (30,50),$ $\beta = 1,$ and $5\%$ significance test. $\texttt{m}$ captures the degree of misspecification in the locations of non-zero network links. The bottom panel displays the average estimate of the peer effect coefficient $\rho$ in TSLS.}
\end{figure}

The top panel of Figure \ref{fig:power3} displays the power of $\texttt{T}_{\texttt{JL}}$ and $\texttt{t-test}_{\texttt{TSLS}}$, and the bottom panel displays the average estimate of the peer effect coefficient $\rho$ in TSLS. The power of $\texttt{t-test}_{\texttt{TSLS}}$ diminishes rapidly as $\texttt{m}$ (i.e., the degree of misspecification) increases, whereas the proposed AR test is robust to the misspecification. The power of $\texttt{t-test}_{\texttt{TSLS}}$ exhibits a nonlinear relationship with the degree of misspecification: it drops below 0.1 at \texttt{m} = 3 and then increases up to 0.4. The average estimate of $\rho$ in TSLS on the bottom panel indicates that the increase in the power is due to the sizable ``negative" estimates after \texttt{m} = 3. It signifies that misspecification can result in not only power loss but also misleading results.  

Lastly, we perform two sets of simulations to examine misspecification in the direction of peer effects.\footnote{We thank an anonymous referee for suggesting this exercise.} Practitioners typically assume homogeneity in the direction of peer effects, but in the following exercises, we allow for the coexistence of positive and negative peer effects. The first exercise adopts the simulation design used for Table \ref{tb:power_ND}, but sets $\alpha_{it} = 0.3$ or $-0.3$, varying the proportion of negative coefficients ($\texttt{prop}_{\texttt{neg}}$). The misspecified adjacency matrix $\mathbf{W}$ for $\texttt{t-t}_{\texttt{TSLS}}$ is obtained by $\mathbf{W}^{\ast}=[w_{ij}^{\ast}]$, where $w_{ij}^{\ast} = \mathbf{1}\{ \alpha_{ij} \ne 0 \}$ for $i \ne j$, and then row-normalize $\mathbf{W}^{\ast}$ to get $\mathbf{W}$. Thus, the size and location of the network links are correctly specified, but the direction of peer interactions is misspecified.
We set $(n,T) = (30,50)$, $\beta = 1$, and $\texttt{ND} = 0.3$. The simulation results are reported in Table \ref{tb:misdirection}.

\begin{table}[H] \singlespacing
\centering
\begin{threeparttable}
\caption{Misspecification in the Direction of Network Effects: Experiment I}
    \label{tb:misdirection}
\begin{tabular}{c|p{2cm}p{2cm}|p{2cm}p{2cm}} \hline \hline
& \multicolumn{2}{c|}{Normal Errors} & \multicolumn{2}{c}{Log-Normal Errors} \\
$\texttt{prop}_{\texttt{neg}}$   & $\texttt{T}_{\texttt{JL}}$    & $\texttt{t-t}_{\texttt{TSLS}}$  & $\texttt{T}_{\texttt{JL}}$    & $\texttt{t-t}_{\texttt{TSLS}}$  \\ \hline
0.1       & 0.980 & 1.000 & 0.987 & 1.000 \\
0.5       & 0.999 & 0.057 & 0.882 & 0.148 \\ \hline \hline
\end{tabular}
\begin{tablenotes}
 \small
 \item Note: $\texttt{prop}_{\texttt{neg}}$ is the proportion of negative peer effect coefficients. $(n,T) = (30,50),$ $\beta = 1,$ $\texttt{ND} = 0.3,$ and $5\%$ significance test. 
  \end{tablenotes}
  \end{threeparttable}
\end{table}

It is clear from the results that the power of $\texttt{T}_{\texttt{JL}}$ is robust to the proportion of negative links, whereas the rejection rate of $\texttt{t-t}_{\texttt{TSLS}}$ significantly decreases as $\texttt{prop}_{\texttt{neg}}$ increases. The TSLS estimate of $\rho$ when $\texttt{prop}_{\texttt{neg}} = 0.5$ is almost zero under normal errors, which explains the deterioration in the power of $\texttt{t-t}_{\texttt{TSLS}}$.

The next simulation uses the design used for Figure \ref{fig:power3}, but assigns a positive weight for one neighbor and a negative weight for the other. Specifically, we consider $n=30$ and divide them into two groups such that the first group includes individuals from 1 to 15 and the other group includes the rest. Each group forms a circular network, and within groups, we set $\alpha_{ij} = 0.3$ if $j$ is $i$'s nearest right-side neighbor. Across groups, we set $\alpha_{ij} = -0.3$ if $j = i + n/2$ for $i \leq n/2$ and $j = i - n/2$ for $i > n/2$. All other coefficients are set to zero. As a result, the peer effects within groups are positive and those across groups are negative. The misspecified adjacency matrix $\mathbf{W}$ for $\texttt{t-t}_{\texttt{TSLS}}$ is obtained from $\mathbf{W}^{\ast}=[w_{ij}^{\ast}]$, where $w_{ij}^{\ast} = \mathbf{1}\{ \alpha_{ij} \ne 0 \}$ for $i \ne j$, and then row-normalize $\mathbf{W}^{\ast}$ to get $\mathbf{W}$. We set $(n,T) = (30,50)$, and  $\beta = 1$. For comparison, we compute the results where the peer effects are positive both within and across groups. The simulation results are reported in Table \ref{misdirection2}.

\begin{table}[H] \singlespacing
\centering
\begin{threeparttable}
\caption{Misspecification in the Direction of Network Effects: Experiment II} \label{misdirection2}
\begin{tabular}{l |p{2cm}p{2cm}|p{2cm}p{2cm}} \hline \hline
& \multicolumn{2}{c|}{Normal Errors} & \multicolumn{2}{c}{Log-Normal Errors} \\
   & $\texttt{T}_{\texttt{JL}}$    & $\texttt{t-t}_{\texttt{TSLS}}$  & $\texttt{T}_{\texttt{JL}}$    & $\texttt{t-t}_{\texttt{TSLS}}$  \\ \hline
Pos. Only    & 0.936 & 1.000 & 0.862 & 1.000 \\
Pos. \& Neg. & 0.956 & 0.250 & 0.893 & 0.250 \\ \hline \hline
\end{tabular}
\begin{tablenotes}
 \small
 \item Note: $(n,T) = (30,50),$ $\beta = 1,$ and $5\%$ significance test. 
  \end{tablenotes}
  \end{threeparttable}
\end{table}

Similar to the previous results, the power of $\texttt{T}_{\texttt{JL}}$ is robust to the direction of peer effects, whereas the rejection rate of $\texttt{t-t}_{\texttt{TSLS}}$ falls below 30\% when both negative and positive peer interactions exist. Overall, these exercises demonstrate that the proposed AR test can be an effective and reliable alternative when the misspecification of the network structure is a concern.

\section{Empirical Applications}\label{emp}

\subsection{Growth Spillovers among OECD Countries}\label{emp1}

We apply our AR test to the international growth spillover model considered in \cite{EK2007} and \cite{HO2013} among others. The papers introduce spatial externalities to the classical Solow growth model by augmenting the model with spatial lags to account for spatial interdependence between countries due to knowledge transfer and technological spillover. The spatially augmented Solow model requires specification of the dependence structure to identify the spatial effects, for which the papers use geographic distance or bilateral trade volume. Since our test does not rely on a particular network structure, our result can be interpreted as more general evidence for global interdependence.

We use a balanced panel of 28 OECD member countries over the period 1975 - 2015 and specify our (unrestricted) model as follows:\footnote{The specification (\ref{model}) is a simplified version of the real income model used in \citet[][equation (23)]{EK2007}. As there is no information about potential peers in the data, we set $%
\mathcal{N}_{i}=\mathcal{N}/\{i\}$. The 28 OECD countries are the countries that joined the OECD by 2010 and have data for the entire period of analysis: Australia, France, Republic of Korea, Sweden, Austria, Greece, Mexico, Switzerland, Belgium, Iceland, Netherlands, Turkey, Canada, Ireland, New Zealand, United Kingdom, Norway, United States, Denmark, Italy, Portugal, Finland, Japan, Spain, Germany, Hungary, Luxembourg, and Poland.} 
\begin{equation}
\ln y_{it}= \sum_{j \in \mathcal{N}/\{i\}} \alpha_{ij}  \ln y_{jt} + \beta_1 \ln (p_{it} + 0.05) + \beta_2 \ln s_{it} + \delta_i + \mu_t + v_{it}.  \label{model}
\end{equation}%
The outcome variable $y_{it}$ is the real GDP per worker. The exogenous variables $p_{it}$ and $s_{it}$ are the average annual working-age population growth and average saving rate, respectively, over the last five years. More specifically, $s_{it}$ is measured by the average investment share in GDP.\footnote{As is common in the literature, we suppose the sum of exogenous technical progress rate and capital depreciation rate in the model is 0.05.} 
We compiled the panel of our analysis from the OECD database (\texttt{https://data.oecd.org}) for the working-age population data and the PennWorld Tables, version 10.0, for the rest of the data.

\begin{table}[h!]
\centering
  \begin{threeparttable}
  \caption{Summary Statistics} \label{tb:3}
\begin{tabular}{c|c|c|c|c|c}
\hline
\hline
     & Mean  & Max   & Median & Min    & SD   \\
     \hline
$y$    & 65,644 & 180,156 & 63,363&13,510  & 25,320 \\
$p$ & 0.009 & 0.033 & 0.007 & -0.011 & 0.008 \\
$s$  & 27.3  & 46.9 &26.5  & 15.1   & 5.7  \\
\hline
\hline
\end{tabular}
 \begin{tablenotes}
 \small
 \item Note: $y$ is the output-side real GDP per worker at chained PPPs (in mil. 2017US\$), $p$ is the annual working-age population growth rate, and $s$ is the share of gross capital formation at current PPPs.
   \end{tablenotes}
    \end{threeparttable}
\end{table}

As discussed in Footnote \ref{ft2} of Section \ref{ARwof}, the IV matrix $\mathbf{Q}$ collects linearly
independent columns in $[\mathbf{X},\mathbf{Z}]$, where $\mathbf{Z}=(\mathbf{Z}_{1}^{\prime },\cdots
,\mathbf{Z}_{T}^{\prime })^{\prime }$ with $\mathbf{Z}_{t}=(\mathbf{e}_{1}\mathbf{Z}_{1t},\cdots ,\mathbf{e}_{n}\mathbf{Z}_{nt})$.
If $\mathbf{Z}_{it}=(\mathbf{X}_{1t},\cdots ,\mathbf{X}_{i-1,t},\mathbf{X}_{i+1,t},\cdots ,\mathbf{X}_{nt})$, then the
total number of IVs is $L+n(n-1)L$, where $L$ is the number of exogenous regressors in $\mathbf{X}_{it}$.
If $L$ is large, the number of IVs could be larger than the sample size and violate the regularity condition that the IV matrix has the full column rank. In this application, to reduce the number of IVs, we use $\mathbf{Z}_{it}=(\mathbf{X}_{1t}\boldsymbol{\iota}
_{L},\cdots ,\mathbf{X}_{i-1,t}\boldsymbol{\iota} _{L},\mathbf{X}_{i+1,t}\boldsymbol{\iota} _{L},\cdots ,\mathbf{X}_{nt}\boldsymbol{\iota}
_{L})$, where $\boldsymbol{\iota} _{L}$ is a $L\times 1$ vector of ones. With IVs constructed in this way, the number of IVs is less
than the sample size as long as $n<T$.

The spatial lag is the source of growth spillovers in this model. Therefore, we test for the presence of growth spillovers by testing $H_0: \alpha_{ij} = 0$ for all $(i,j)$. Our AR test strongly supports the presence of global spillovers with a near-zero p-value.\footnote{The test statistic for the chi-square approximation is 1070.70, while the critical value for the nominal 1\% test is 810.22.} Our result echoes the significant spillover effects that have been identified in \citet{EK2007} and \citet{HO2013}. However, compared to the existing studies, our result does not rely on any specification assumption for the underlying network structure and is thus more robust.

\subsection{Player Interaction in the NBA}\label{emp2}
Our second application examines player interactions in the National Basketball Association (NBA) games. We use the NBA 2015-16 season data used in  \citet{HJS20},\footnote{They estimate peer effects among NBA players but their empirical model imposes a particular network structure such that players are affected only by the same type of players, where ``types" are the player positions: $Guards$ or $Forwards$.} and follow the paper to create the outcome and exogenous variables for our empirical model. The data include player-period level offensive and defensive statistics such as points, fouls, steals and etc, where a period represents any contiguous game period in which the same ten players are on the court. In this case, the player networks are time-varying, so the data are conceptualized for repeated cross-sections. 

As our model requires a panel, we focus on the most frequently used lineups of players in the eastern and western conference winners of the season, Cleveland (CLE) and Golden State Warriors (GSW), respectively, and construct panels for the two lineups for the season. The lineup for CLE includes L. James, K. Love, J. Smith, T. Thompson, and K. Irving, and the lineup for GSW includes H. Barnes, D. Green, A. Bogut, K. Thompson, and S. Curry. The panel for CLE (GSW) includes 140 (106) time periods and spans 306.1 (307) minutes in total. Hereafter, we call the two lineups the best lineups.

The empirical model is similar to Equation (\ref{model}) and uses the \emph{Wins Produced} for the outcome variable, which is a leading measure of NBA player production based on the work of sports economist \citet{Berri99}:
\begin{equation} \begin{aligned} \nonumber
y_{it} = &(0.064 \cdot 3PT_{it} + 0.032 \cdot 2PT_{it} + 0.017 \cdot FT_{it} + 0.034 \cdot REB_{it} + 0.033 \cdot STL_{it}  \\
&+ 0.020 \cdot BLK_{it} - 0.034 \cdot MFG_{it} - 0.015 \cdot MFT_{it} - 0.034 \cdot TO_{it})/Mins_{it},
\end{aligned} \end{equation} 
where $3PT_{it}$, $2PT_{it}$, $FT_{it}$, $REB_{it}$, $STL_{it}$, $BLK_{it}$, $MFG_{it}$, $TO_{it}$, and $Mins_{it}$ are 3-point field goals made, 2-point field goals made, free throws made, rebounds, steals, blocks, missed field goals, missed free throws, turnovers, and minutes played, respectively, by player $i$ in period $t$. \emph{Wins Produced} per minute (or wins per minute) estimates a player's marginal win productivity based upon player-level variables related to team-winning. 

Our exogenous variables include two player-level exogenous variables, $Experience_{it}$ and $Fatigue_{it}$.\footnote{ \citet{HJS20} also include three ``team-level" exogenous variables, which are controlled for by the time fixed effect in our model.} The $Experience_{it}$ is minutes played from the start of the game to the end of period $t-1$, and $Fatigue_{it}$ is minutes continuously played until the end of period $t-1$.

\begin{table}[h!]
\caption{AR Test for NBA Player Interactions} \label{tb:4}
\centering
\begin{tabular}{l|cc}
\hline \hline
                  & \textbf{CLE}     & \textbf{GSW}       \\
                  \hline
Test statistic         & 44.20  & 24.39  \\
P-value & 0.23   & 0.96  \\ \hline \hline
\end{tabular}
\end{table} 

Table \ref{tb:4} includes AR test statistics for player interaction in the best lineups over the season, where the statistics are computed for the chi-square distribution approximation. All test statistics fail to reject the null at 5\% significance level, which implies that interactions among players in the two lineups were not significant. Overall, it appears that the data contains little signal after the two-way transformation, rendering very small estimated coefficients for the exogenous variables under the null. According to the simulation results, this is the case where the power of our test can be weak. 

\begin{figure}[ht!]
	\caption{Rolling Window P-Values for the Best Lineups Over 2015-16 Season}
		 \label{fig:4} 
	\includegraphics[width=1\linewidth, trim={2.5cm 0cm 2.5cm 1.5cm},clip]{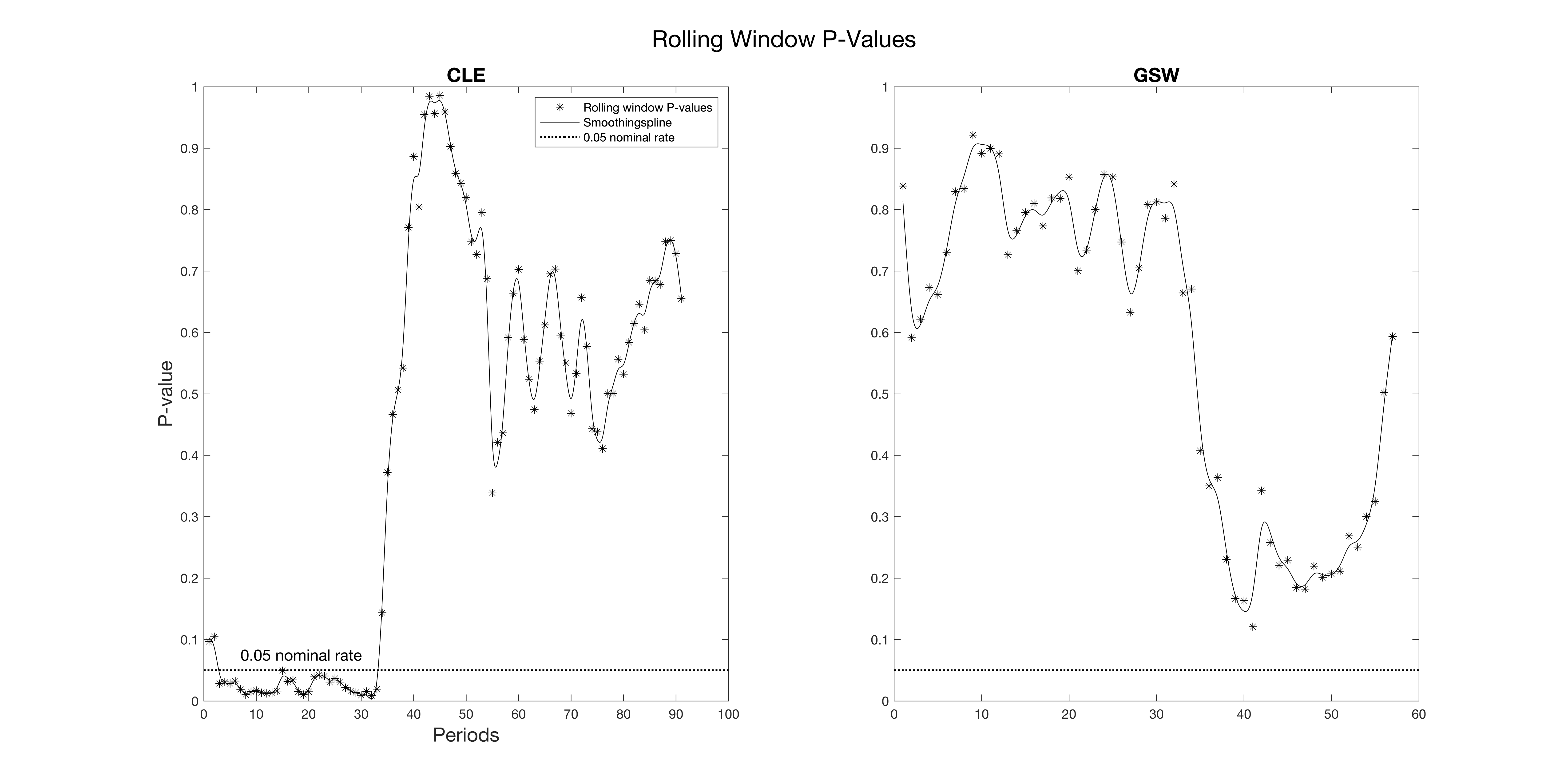}
\end{figure}

However, interactions between players may vary over time and the test statistics aggregated for the entire season may mask the time-varying interaction effects. Therefore, we also examine the changes in the test statistic over the season using a rolling window analysis, where we repeatedly compute the p-value of the AR test statistic with a rolling window of fifty time periods. Figure \ref{fig:4} plots the p-values for the best lineups over the season.\footnote{The splines are computed using the function ``fit" with ``smoothingspline" option in MatLab. The smoothing parameter was automatically determined by the function.}

In the case of GSW, p-values are still quite large. However, in the case of CLE, the p-value falls significantly below 0.05 between the 5th and 30th time periods, indicating that the player performances in the lineup were significantly interdependent during that time period. Overall, we observe substantial heterogeneity in the test statistics for the two lineups over the season, which strongly suggests that player interactions are not fixed, but change over time to adapt to the game environment. 

\section{Conclusion}\label{con}
This paper proposes an AR test for the statistical significance of dyad-specific peer effects in a linear panel data model of social interactions. The main advantage of the proposed test is that it does not require specifying the interaction structure. In our test, both the number of null restrictions and the number of IVs employed to test them increase with the sample size, bridging the literature on network models with unknown interaction structures and the literature on inference with many restrictions and/or IVs. 

An important assumption of the proposed test is that network effects are invariant over time. When the true network effects are time-varying, the proposed AR test may yield misleading results. A partial solution is to conduct a rolling window analysis as illustrated in Section \ref{emp2}. Another restrictive assumption is that the random shocks $\epsilon_{it}$ need to be i.i.d. in the presence of individual and time fixed effects. The jackknife AR test statistic presented in Section \ref{ARwof} is robust to heteroskedasticity of unknown form, but the standard jackknife method does not properly re-center the test statistic in the presence of two-way fixed effects. Although including fixed effects can alleviate potential heteroskedasticity to a certain extent, developing a heteroskedasticity-robust test for network effects in the presence of two-way fixed effects remains an important direction 
for future research.\\

\newpage
\appendix{}

\section{Proofs}

\begin{proof}[\textbf{Proof of Proposition \protect\ref{prop1}}]
Under $H_{0}:\boldsymbol{\alpha} =\mathbf{0}$,%
\begin{eqnarray*}
&&\frac{1}{\sqrt{K}}\widetilde{\mathbf{u}}^{\prime }(\mathbf{P}-\mathbf{D})\widetilde{\mathbf{u}} \\
&=&\frac{1}{\sqrt{K}}[\mathbf{u}-\mathbf{X}(\widetilde{\boldsymbol{\beta} }-\boldsymbol{\beta} )]^{\prime }(\mathbf{P}-\mathbf{D})[\mathbf{u}-\mathbf{X}(%
\widetilde{\boldsymbol{\beta} }-\boldsymbol{\beta} )] \\
&=&\frac{1}{\sqrt{K}}\mathbf{u}^{\prime }(\mathbf{P}-\mathbf{D})\mathbf{u}+\frac{1}{\sqrt{K}}(\widetilde{\boldsymbol{\beta} }%
-\boldsymbol{\beta} )^{\prime }\mathbf{X}^{\prime }(\mathbf{P}-\mathbf{D})\mathbf{X}(\widetilde{\boldsymbol{\beta} }-\boldsymbol{\beta} ) \\
&&-\frac{1}{\sqrt{K}}(\widetilde{\boldsymbol{\beta} }-\boldsymbol{\beta} )^{\prime }\mathbf{X}^{\prime }(\mathbf{P}-\mathbf{D})\mathbf{u}-%
\frac{1}{\sqrt{K}}\mathbf{u}^{\prime }(\mathbf{P}-\mathbf{D})\mathbf{X}(\widetilde{\boldsymbol{\beta} }-\boldsymbol{\beta} ).
\end{eqnarray*}%
As $\mathbf{PX}=\mathbf{X}$, it follows by Assumption 3 that 
\begin{equation*}
\frac{1}{\sqrt{K}}(\widetilde{\boldsymbol{\beta} }-\boldsymbol{\beta} )^{\prime }\mathbf{X}^{\prime }(\mathbf{P}-\mathbf{D})\mathbf{X}(%
\widetilde{\boldsymbol{\beta} }-\boldsymbol{\beta} )=\frac{1}{\sqrt{K}}\sqrt{N}(\widetilde{\boldsymbol{\beta} }%
-\boldsymbol{\beta} )^{\prime }\frac{1}{N}(\mathbf{X}^{\prime }\mathbf{X}-\mathbf{X}^{\prime }\mathbf{DX})\sqrt{N}(\widetilde{%
\boldsymbol{\beta} }-\boldsymbol{\beta} )=o_{p}(1).
\end{equation*}%
Similarly,%
\begin{equation*}
\frac{1}{\sqrt{K}}(\widetilde{\boldsymbol{\beta} }-\boldsymbol{\beta} )^{\prime }\mathbf{X}^{\prime }(\mathbf{P}-\mathbf{D})\mathbf{u}=%
\frac{1}{\sqrt{K}}\sqrt{N}(\widetilde{\boldsymbol{\beta} }-\boldsymbol{\beta} )^{\prime }\frac{1}{%
\sqrt{N}}(\mathbf{X}^{\prime }\mathbf{u}-\mathbf{X}^{\prime }\mathbf{Du}),
\end{equation*}%
where $\mathrm{Var}(\frac{1}{\sqrt{N}}\mathbf{X}^{\prime }\mathbf{u}|\mathbf{X})=\frac{1}{N}\mathbf{X}^{\prime
}\boldsymbol{\Omega} \mathbf{X}$ and $\mathrm{Var}(\frac{1}{\sqrt{N}}\mathbf{X}^{\prime }\mathbf{Du}|\mathbf{X})=\frac{1}{N}%
\mathbf{X}^{\prime }\mathbf{D}\boldsymbol{\Omega} \mathbf{DX}$, and, hence, it follows by Assumption 3 and Markov's
inequality that 
\begin{equation*}
\frac{1}{\sqrt{K}}(\widetilde{\boldsymbol{\beta} }-\boldsymbol{\beta} )^{\prime }\mathbf{X}^{\prime
}(\mathbf{P}-\mathbf{D})\mathbf{u}=o_{p}(1).
\end{equation*}%
Finally, as $\varsigma _{it}^{2}\geq \underline{\varsigma }^{2}>0$ and $%
p_{ii}\leq C_{p}<1$,%
\begin{eqnarray*}
\Phi &=&\frac{2}{K}\mathrm{tr}[(\mathbf{P}-\mathbf{D})\boldsymbol{\Omega} (\mathbf{P}-\mathbf{D})\boldsymbol{\Omega} ]\geq \frac{2%
\underline{\varsigma }^{4}}{K}\mathrm{tr}[(\mathbf{P}-\mathbf{D})^{2}]=\frac{2\underline{%
\varsigma }^{4}}{K}\mathrm{tr}(\mathbf{P}^{2}-\mathbf{DP}-\mathbf{PD}+\mathbf{D}^{2}) \\
&=&\frac{2\underline{\varsigma }^{4}}{K}[\mathrm{tr}(\mathbf{P})-\mathrm{tr}%
(\mathbf{D}^{2})]\geq 2\underline{\varsigma }^{4}[1-\frac{C_{p}}{K}\mathrm{tr}(\mathbf{D})]=2%
\underline{\varsigma }^{4}(1-C_{p})>0.
\end{eqnarray*}%
Hence, it follows by Lemma A2 of \citet{Chao2012} that 
\begin{equation*}
\frac{1}{\sqrt{K}\sqrt{\Phi }}\mathbf{u}^{\prime }(\mathbf{P}-\mathbf{D})\mathbf{u}\overset{d}{\rightarrow }%
N(0,1).
\end{equation*}%
The desired result follows as $\widetilde{\Phi }-\Phi =o_{p}(1)$.
\end{proof}

\begin{proof}[\textbf{Proof of Proposition \protect\ref{prop2}}]
Let $\lambda ^{\ast }=K^{\ast }/N^{\ast }$. Under $H_{0}:\boldsymbol{\alpha} =\mathbf{0}$, as $%
\mathbf{P}^{\ast }\mathbf{X}^{\ast }=\mathbf{X}^{\ast }$, $\boldsymbol{\epsilon} ^{\ast \prime }\boldsymbol{\epsilon} ^{\ast
}=\boldsymbol{\epsilon} ^{\prime }\mathbf{J}\boldsymbol{\epsilon} $, and $\boldsymbol{\epsilon} ^{\ast \prime }\mathbf{P}^{\ast
}\boldsymbol{\epsilon} ^{\ast }=\boldsymbol{\epsilon} ^{\prime }\mathbf{P}^{\ast }\boldsymbol{\epsilon} $, we have%
\begin{eqnarray*}
&&\frac{1}{\sqrt{K^{\ast }}}\widehat{\boldsymbol{\epsilon} }^{\ast \prime }(\mathbf{P}^{\ast
}-\lambda ^{\ast }\mathbf{I}_{N})\widehat{\boldsymbol{\epsilon} }^{\ast } \\
&=&\frac{1}{\sqrt{K^{\ast }}}[\boldsymbol{\epsilon} ^{\ast }-\mathbf{X}^{\ast }(\widehat{\boldsymbol{\beta} }%
-\boldsymbol{\beta} )]^{\prime }(\mathbf{P}^{\ast }-\lambda ^{\ast }\mathbf{I}_{N})[\boldsymbol{\epsilon} ^{\ast
}-\mathbf{X}^{\ast }(\widehat{\boldsymbol{\beta} }-\boldsymbol{\beta} )] \\
&=&\frac{1}{\sqrt{K^{\ast }}}\boldsymbol{\epsilon} ^{\prime }(\mathbf{P}^{\ast }-\lambda ^{\ast
}\mathbf{J})\boldsymbol{\epsilon} +\frac{1}{\sqrt{K^{\ast }}}(1-\lambda ^{\ast })(\widehat{\boldsymbol{\beta} }%
-\boldsymbol{\beta} )^{\prime }\mathbf{X}^{\ast \prime }\mathbf{X}^{\ast }(\widehat{\boldsymbol{\beta} }-\boldsymbol{\beta} ) \\
&&-\frac{1}{\sqrt{K^{\ast }}}(1-\lambda ^{\ast })(\widehat{\boldsymbol{\beta} }-\boldsymbol{\beta}
)^{\prime }\mathbf{X}^{\ast \prime }\boldsymbol{\epsilon} ^{\ast }-\frac{1}{\sqrt{K^{\ast }}}%
(1-\lambda ^{\ast })\boldsymbol{\epsilon} ^{\ast \prime }\mathbf{X}^{\ast }(\widehat{\boldsymbol{\beta} }-\boldsymbol{\beta}
) \\
&=&\frac{1}{\sqrt{K^{\ast }}}\boldsymbol{\epsilon} ^{\prime }(\mathbf{P}^{\ast }-\lambda ^{\ast
}\mathbf{J})\boldsymbol{\epsilon} +o_{p}(1),
\end{eqnarray*}%
where the last equality follows by Assumptions 1'-3' and Markov's
inequality. As 
\begin{equation*}
\mathbf{J}=(\mathbf{I}_{T}-\frac{1}{T}\boldsymbol{\iota} _{T}\boldsymbol{\iota} _{T}^{\prime })\otimes (\mathbf{I}_{n}-\frac{1}{n}%
\boldsymbol{\iota} _{n}\boldsymbol{\iota} _{n}^{\prime })=\mathbf{I}_{N}-\frac{1}{T}(\boldsymbol{\iota} _{T}\boldsymbol{\iota}
_{T}^{\prime })\otimes \mathbf{I}_{n}-\frac{1}{n}\mathbf{I}_{T}\otimes (\boldsymbol{\iota} _{n}\boldsymbol{\iota}
_{n}^{\prime })+\frac{1}{N}(\boldsymbol{\iota} _{T}\boldsymbol{\iota} _{T}^{\prime })\otimes (\boldsymbol{\iota}
_{n}\boldsymbol{\iota} _{n}^{\prime }),
\end{equation*}%
where $N=nT$, we have%
\begin{eqnarray*}
&&\frac{1}{\sqrt{K^{\ast }}}\boldsymbol{\epsilon} ^{\prime }(\mathbf{P}^{\ast }-\lambda ^{\ast
}\mathbf{J})\boldsymbol{\epsilon}  \\
&=&\frac{1}{\sqrt{K^{\ast }}}\boldsymbol{\epsilon} ^{\prime }[\mathbf{P}^{\ast }-\lambda ^{\ast
}\mathbf{I}_{N}+\lambda ^{\ast }\frac{1}{T}(\boldsymbol{\iota} _{T}\boldsymbol{\iota} _{T}^{\prime })\otimes
\mathbf{I}_{n}+\lambda ^{\ast }\frac{1}{n}\mathbf{I}_{T}\otimes (\boldsymbol{\iota} _{n}\boldsymbol{\iota} _{n}^{\prime
})-\lambda ^{\ast }\frac{1}{N}(\boldsymbol{\iota} _{T}\boldsymbol{\iota} _{T}^{\prime })\otimes
(\boldsymbol{\iota} _{n}\boldsymbol{\iota} _{n}^{\prime })]\boldsymbol{\epsilon}  \\
&=&\frac{1}{\sqrt{K^{\ast }}}\boldsymbol{\epsilon} ^{\prime }(\mathbf{P}^{\ast }-\lambda
\mathbf{I}_{N})\boldsymbol{\epsilon} +\frac{\lambda ^{\ast }}{\sqrt{K^{\ast }}}\boldsymbol{\epsilon} ^{\prime
}\mathbf{M}\boldsymbol{\epsilon} ,
\end{eqnarray*}%
with $\lambda =K^{\ast }/N$ and%
\begin{equation*}
\mathbf{M}=\frac{\lambda -\lambda ^{\ast }}{\lambda ^{\ast }}\mathbf{I}_{N}+\frac{1}{T}(\boldsymbol{\iota}
_{T}\boldsymbol{\iota} _{T}^{\prime })\otimes \mathbf{I}_{n}+\frac{1}{n}\mathbf{I}_{T}\otimes (\boldsymbol{\iota}
_{n}\boldsymbol{\iota} _{n}^{\prime })-\frac{1}{N}(\boldsymbol{\iota} _{T}\boldsymbol{\iota} _{T}^{\prime
})\otimes (\boldsymbol{\iota} _{n}\boldsymbol{\iota} _{n}^{\prime }).
\end{equation*}%
As $\mathbf{M}$ is symmetric with a zero diagonal, $\mathrm{E}(\boldsymbol{\epsilon} ^{\prime
}\mathbf{M}\boldsymbol{\epsilon} )=0$ and 
\begin{equation*}
\mathrm{Var}(\boldsymbol{\epsilon} ^{\prime }\mathbf{M}\boldsymbol{\epsilon} )=2\sigma ^{4}\mathrm{tr}%
(\mathbf{M}^{2})=2\sigma ^{4}(n+T-\frac{n}{T}-\frac{T}{n}+\frac{2}{T}+\frac{2}{n}-%
\frac{1}{N}-3).
\end{equation*}%
If $\lambda ^{\ast }\rightarrow c$ for some $0<c<1$, then $K^{\ast }$ goes
to infinity at the same rate as $N^{\ast }$, which implies $\mathrm{Var}(%
\frac{\lambda ^{\ast }}{\sqrt{K^{\ast }}}\boldsymbol{\epsilon} ^{\prime }\mathbf{M}\boldsymbol{\epsilon} )=%
\frac{(\lambda ^{\ast })^{2}}{K^{\ast }}\mathrm{Var}(\boldsymbol{\epsilon} ^{\prime
}\mathbf{M}\boldsymbol{\epsilon} )=o(1)$. On the other hand, if $\lambda ^{\ast }\rightarrow 0$,
then $\mathrm{Var}(\frac{\lambda ^{\ast }}{\sqrt{K^{\ast }}}\boldsymbol{\epsilon}
^{\prime }\mathbf{M}\boldsymbol{\epsilon} )=\frac{(\lambda ^{\ast })^{2}}{K^{\ast }}\mathrm{Var}%
(\boldsymbol{\epsilon} ^{\prime }\mathbf{M}\boldsymbol{\epsilon} )=\frac{\lambda ^{\ast }}{N^{\ast }}\mathrm{Var%
}(\boldsymbol{\epsilon} ^{\prime }\mathbf{M}\boldsymbol{\epsilon} )=o(1)$. Hence, for both cases, $\frac{%
\lambda ^{\ast }}{\sqrt{K^{\ast }}}\boldsymbol{\epsilon} ^{\prime }\mathbf{M}\boldsymbol{\epsilon} =o_{p}(1)$
by Markov's inequality, which implies 
\begin{equation*}
\frac{1}{\sqrt{K^{\ast }}}\boldsymbol{\epsilon} ^{\prime }(\mathbf{P}^{\ast }-\lambda ^{\ast
}\mathbf{J})\boldsymbol{\epsilon} =\frac{1}{\sqrt{K^{\ast }}}\boldsymbol{\epsilon} ^{\prime }(\mathbf{P}^{\ast }-\lambda
\mathbf{I}_{N})\boldsymbol{\epsilon} +o_{p}(1).
\end{equation*}%
With a little abuse of the notation, let $\epsilon _{i}$ denote the $i$th
element of $\boldsymbol{\epsilon} $ and $p_{ij}^{\ast }$ denote the $(i,j)$th element of $%
\mathbf{P}^{\ast }$. As $\sum_{i}(p_{ii}^{\ast }-\lambda )=0$,%
\begin{eqnarray*}
&&\frac{1}{\sqrt{K^{\ast }}}\boldsymbol{\epsilon} ^{\prime }(\mathbf{P}^{\ast }-\lambda
\mathbf{I}_{N})\boldsymbol{\epsilon}  \\
&=&\frac{1}{\sqrt{K^{\ast }}}\sum_{i}\sum_{j\neq i}p_{ij}^{\ast }\epsilon
_{i}\epsilon _{j}+\frac{1}{\sqrt{K^{\ast }}}\sum_{i}(p_{ii}^{\ast }-\lambda
)\epsilon _{i}^{2}-\frac{\sigma ^{2}}{\sqrt{K^{\ast }}}\sum_{i}(p_{ii}^{\ast
}-\lambda ) \\
&=&\frac{1}{\sqrt{K^{\ast }}}\sum_{i}\sum_{j\neq i}p_{ij}^{\ast }\epsilon
_{i}\epsilon _{j}+\sum_{i}\omega _{i},
\end{eqnarray*}%
where $\omega _{i}=(p_{ii}^{\ast }-\lambda )(\epsilon _{i}^{2}-\sigma ^{2})/%
\sqrt{K^{\ast }}$. By construction, $\mathrm{E}(\omega _{i}|\mathbf{X}_{i})=0$,
where, with a little abuse of the notation, $\mathbf{X}_{i}$ denotes the $i$th row of 
$\mathbf{X}$. Under Assumptions 1' and 2', $\sum_{i}\mathrm{E}(\omega_{i}^{4})\rightarrow 0$ since $|p_{ii}^{\ast }-\lambda | \leq \text{max}\{p_{ii}^{\ast },\lambda \}$, and $\sum_{i}(p_{ii}^{\ast })^{4} < \sum_{i}p_{ii}^{\ast } = K^{\ast}$ when $K^{\ast}/N < 1$. Hence, it
follows by Lemma A2 of \citet{Hansen2008} that 
\begin{equation*}
\frac{1}{\sqrt{K^{\ast }}\sqrt{\Phi ^{\ast }}}\boldsymbol{\epsilon} ^{\prime }(\mathbf{P}^{\ast
}-\lambda \mathbf{I}_{N})\boldsymbol{\epsilon} \overset{d}{\rightarrow }N(0,1),
\end{equation*}%
where%
\begin{equation*}
\Phi ^{\ast }=\mathrm{E}[(\epsilon _{i}^{2}-\sigma ^{2})^{2}]\mathrm{plim}%
_{N\rightarrow \infty }\frac{1}{K^{\ast }}\sum_{i}(p_{ii}^{\ast
}-\lambda )^{2}+2\sigma ^{4}[1-\mathrm{plim}_{N\rightarrow \infty }%
\frac{1}{K^{\ast }}\sum_{i}(p_{ii}^{\ast })^{2}].
\end{equation*}%
Since $\sum_{i}(p_{ii}^{\ast })^{2} < \sum_{i}p_{ii}^{\ast } = K^{\ast }$ when $K^{\ast}/N < 1$,
we have $\mathrm{plim}_{N\rightarrow \infty }\frac{1}{K^{\ast }}%
\sum_{i}(p_{ii}^{\ast })^{2} < 1$, which implies $\Phi ^{\ast }>0$. 
Rearranging terms, we can write $\Phi ^{\ast }$ as%
\begin{equation*}
\Phi ^{\ast }=(\mu _{4}-3\sigma ^{4})[\mathrm{plim}_{N\rightarrow
\infty }\frac{1}{K^{\ast }}\sum_{i}(p_{ii}^{\ast })^{2}-\bar{\lambda}%
]+2\sigma ^{4}(1-\bar{\lambda}),
\end{equation*}%
where $\bar{\lambda}=\lim_{N\rightarrow \infty }K^{\ast }/N$. The
desired result follows as $\widehat{\Phi }^{\ast }-\Phi ^{\ast }=o_{p}(1)$.
\end{proof}

\begin{proof}[\textbf{Proof of Corollary}] (i) When $K^{\ast} \rightarrow \infty$: Since $\sqrt{2K^{\ast }}AR_{FE} + K^{\ast } \geq q_{K^{\ast } - L} \left( 1 - \tau \right)$ $\equiv$ $ AR_{FE} \geq \left( q_{K^{\ast } - L} \left( 1 - \tau \right) - K^{\ast } \right)/\sqrt{2K^{\ast }}$, the desired result follows from Proposition \protect\ref{prop2} and that
\begin{equation*}
\frac{q_{K^{\ast } - L} \left( 1 - \tau \right) - K^{\ast }}{\sqrt{2K^{\ast }}} = \sqrt{\frac{K^{\ast } - L}{K^{\ast } }} \left( \frac{q_{K^{\ast } - L} \left( 1 - \tau \right) - \left(K^{\ast } - L \right) }{\sqrt{2\left(K^{\ast } - L \right)}}\right) -  \frac{L}{\sqrt{2K^{\ast }}} \longrightarrow q(1-\tau)
\end{equation*}
as $K^{\ast} \rightarrow \infty$, where $q(1-\tau)$ is the $(1-\tau)$th quantile of the standard normal distribution and the convergence is due to the well-known property of chi-square distribution such that $ \left(q_{f}(1-\tau) - f \right)/\sqrt{2f}$ $ \longrightarrow q(1-\tau)$ as $f \to \infty$.

(ii) When $K^{\ast}$ is fixed: Notice that $\Phi^{\ast} \rightarrow 2\sigma^4$ as $N \rightarrow \infty$. Similarly as in the proof of Proposition \ref{prop2}, we have 
\begin{eqnarray*}
&& \widehat{\boldsymbol{\epsilon} }^{\ast \prime }(\mathbf{P}^{\ast
}-\lambda ^{\ast }\mathbf{I}_{N})\widehat{\boldsymbol{\epsilon} }^{\ast } \\
&=&\boldsymbol{\epsilon} ^{\prime }(\mathbf{P}^{\ast }-\lambda ^{\ast
}\mathbf{J})\boldsymbol{\epsilon} +(1-\lambda ^{\ast })(\widehat{\boldsymbol{\beta} }%
-\boldsymbol{\beta} )^{\prime }\mathbf{X}^{\ast \prime }\mathbf{X}^{\ast }(\widehat{\boldsymbol{\beta} }-\boldsymbol{\beta} ) \\
&&-(1-\lambda ^{\ast })(\widehat{\boldsymbol{\beta} }-\boldsymbol{\beta}
)^{\prime }\mathbf{X}^{\ast \prime }\boldsymbol{\epsilon} ^{\ast }-
(1-\lambda ^{\ast })\boldsymbol{\epsilon} ^{\ast \prime }\mathbf{X}^{\ast }(\widehat{\boldsymbol{\beta} }-\boldsymbol{\beta}
) \\
&=& \boldsymbol{\epsilon} ^{\prime }(\mathbf{P}^{\ast }-\lambda ^{\ast
}\mathbf{J})\boldsymbol{\epsilon} - (1-\lambda ^{\ast }) \boldsymbol{\epsilon}^{\prime} \mathbf{X}^{\ast } (\mathbf{X}^{\ast \prime } \mathbf{X}^{\ast})^{-1 }\mathbf{X}^{\ast \prime }\boldsymbol{\epsilon} \\
&=& \boldsymbol{\epsilon} ^{\prime }(\mathbf{P}^{\ast }-\lambda \mathbf{I}_{N})\boldsymbol{\epsilon} +\lambda ^{\ast }\boldsymbol{\epsilon} ^{\prime}\mathbf{M}\boldsymbol{\epsilon} - (1- \lambda^{\ast})\boldsymbol{\epsilon} ^{\prime }\mathbf{P}_{\mathbf{X}^{\ast}}\boldsymbol{\epsilon}\\
&=& \boldsymbol{\epsilon} ^{\prime }(\mathbf{P}^{\ast }- \mathbf{P}_{\mathbf{X}^{\ast}} - \lambda \mathbf{I}_{N})\boldsymbol{\epsilon} +\lambda ^{\ast }\boldsymbol{\epsilon} ^{\prime}\mathbf{M}\boldsymbol{\epsilon} + \lambda^{\ast}\boldsymbol{\epsilon} ^{\prime }\mathbf{P}_{\mathbf{X}^{\ast}}\boldsymbol{\epsilon},
\end{eqnarray*}%
where $\mathbf{P}_{\mathbf{X}^{\ast}} = \mathbf{X}^{\ast } (\mathbf{X}^{\ast \prime } \mathbf{X}^{\ast})^{-1 }\mathbf{X}^{\ast \prime }$. Using the arguments in the proof of Proposition \ref{prop2}, we have $\lambda ^{\ast }\boldsymbol{\epsilon} ^{\prime}\mathbf{M}\boldsymbol{\epsilon} = o_{p}(1)$. Also, when $K^{\ast}$ is fixed, as $N \rightarrow \infty$, $\mathrm{E}(\lambda^{\ast}\boldsymbol{\epsilon} ^{\prime
}\mathbf{P}_{\mathbf{X}^{\ast}}\boldsymbol{\epsilon} )=\lambda^{\ast} L \rightarrow 0$, and 
\begin{equation*}
\mathrm{Var}(\lambda^{\ast}\boldsymbol{\epsilon} ^{\prime
}\mathbf{P}_{\mathbf{X}^{\ast}}\boldsymbol{\epsilon})= (\lambda^{\ast})^2 \left[ (\mu_4 - 3\sigma^4)\sum_{i}(\mathbf{P}_{\mathbf{X}^{\ast},ii})^2 + 2\sigma ^{4}\mathrm{tr}%
(\mathbf{P}_{\mathbf{X}^{\ast}}) \right] \rightarrow 0,
\end{equation*}%
because $\sum_{i}(\mathbf{P}_{\mathbf{X}^{\ast},ii})^2 \leq \sum_{i}\mathbf{P}_{\mathbf{X}^{\ast},ii} = \mathrm{tr}%
(\mathbf{P}_{\mathbf{X}^{\ast}}) = L,$ 
where, with a little abuse of the notation, $\mathbf{P}_{\mathbf{X}^{\ast},ii}$ denotes the $(i,i)$th element of 
$\mathbf{P}_{\mathbf{X}^{\ast}}$. Therefore, $ \lambda^{\ast}\boldsymbol{\epsilon} ^{\prime }\mathbf{P}^{\ast}_{\mathbf{X}}\boldsymbol{\epsilon} = o_{p}(1).$ Furthermore, $\mathrm{E}(\lambda \boldsymbol{\epsilon} ^{\prime }\boldsymbol{\epsilon})= K^{\ast}\sigma^2$ and
\begin{equation*}
\mathrm{Var}(\lambda \boldsymbol{\epsilon} ^{\prime }\boldsymbol{\epsilon})= \lambda K^{\ast}(\mu_4 - \sigma^4)  \rightarrow 0,
\end{equation*}%
which implies $\lambda \boldsymbol{\epsilon} ^{\prime }\boldsymbol{\epsilon} - K^{\ast}\sigma^2 = o_{p}(1).$
Combining the results above gives
\begin{equation*}
\sqrt{2K^{\ast}}AR_{FE} + K^{\ast} = \boldsymbol{\epsilon} ^{\prime }(\mathbf{P}^{\ast }- \mathbf{P}_{\mathbf{X}^{\ast}})\boldsymbol{\epsilon}/\sigma^2 + o_{p}(1).
\end{equation*}%
Let $\mathbf{Q}^{\circ} = \mathbf{Q}^{\ast}(\mathbf{Q}^{\ast \prime} \mathbf{Q}^{\ast})^{-1/2}$ and $\mathbf{P}_{\mathbf{Q}^{\circ \prime}\mathbf{X}^{\ast}} = \mathbf{Q}^{\circ \prime}\mathbf{X}^{\ast}(\mathbf{X}^{\ast \prime}\mathbf{Q}^{\circ}\mathbf{Q}^{\circ'}\mathbf{X}^{\ast})^{-1}\mathbf{X}^{\ast \prime}\mathbf{Q}^{\circ}.$ Then, $\mathbf{P}^{\ast }- \mathbf{P}_{\mathbf{X}^{\ast}} = \mathbf{P}^{\ast }- \mathbf{P}^{\ast }\mathbf{X}^{\ast } (\mathbf{X}^{\ast \prime }\mathbf{P}^{\ast } \mathbf{X}^{\ast})^{-1 }\mathbf{X}^{\ast \prime }\mathbf{P}^{\ast } = \mathbf{Q}^{\circ} (\mathbf{I}_{K^{\ast}} - \mathbf{P}_{\mathbf{Q}^{\circ \prime}\mathbf{X}^{\ast}})\mathbf{Q}^{\circ \prime},$ which implies $\boldsymbol{\epsilon} ^{\prime }(\mathbf{P}^{\ast }- \mathbf{P}^{\ast}_{\mathbf{X}})\boldsymbol{\epsilon}/\sigma^2 = \boldsymbol{\epsilon} ^{\prime }\mathbf{Q}^{\circ} (\mathbf{I}_{K^{\ast}} - \mathbf{P}_{\mathbf{Q}^{\circ \prime}\mathbf{X}^{\ast}})\mathbf{Q}^{\circ \prime}\boldsymbol{\epsilon}/\sigma^2 \overset{d}{\rightarrow} \chi^2 (K^{\ast}-L)$ by the stated assumptions and the standard arguments about quadratic forms of normal random variables.
\end{proof}
\numberwithin{equation}{section}

\section{Power Analysis}\label{App_Power_Analysis}

Under the alternative described in Remark \ref{power}, the outcome model in period $t$ is given as%
\begin{equation*} 
\mathbf{y}_t=\mathbf{A}_n\mathbf{y}_t+\beta\mathbf{x}_t + \mathbf{u}_t,
\end{equation*}%
where, for simplicity, we assume that there is a single exogenous regressor $\mathbf{x}_t$ in the model and the coefficient $\beta$ is known.\footnote{In the remark at the end of this appendix, we discuss how the estimation of $\beta$ affects the power of the test.} These simplifications do not alter the main message of the analysis.

For the existence of the reduced form, we assume $|\mathbf{I}_{n} - \mathbf{A}_n| \ne 0$, which requires $\prod_{i=1}^n (1 - \tau_i ) \ne 0$ where $\tau_i$'s are the eigenvalues of $\mathbf{A}_n$. Then, the reduced form of the model in period $t$ is given as
\begin{eqnarray*}
\mathbf{y}_t&=&(\mathbf{I}_{n} - \mathbf{A}_n)^{-1}(\beta\mathbf{x}_t +\mathbf{u}_t)\\
& = & \beta\mathbf{x}_t +\beta\mathbf{A}_n(\mathbf{I}_{n} - \mathbf{A}_n)^{-1}\mathbf{x}_t + (\mathbf{I}_{n} - \mathbf{A}_n)^{-1}\mathbf{u}_t.
\end{eqnarray*}

Under the simplifying assumption that $\beta$ is known, the residual vector in period $t$ is given as
\begin{equation}
\widetilde{\mathbf{u}}_t=\mathbf{y}_t - \beta\mathbf{x}_t
 =   \underbrace{\beta\mathbf{A}_n(\mathbf{I}_{n} - \mathbf{A}_n)^{-1}\mathbf{x}_t}_{**} + (\mathbf{I}_{n} - \mathbf{A}_n)^{-1}\mathbf{u}_t, \label{eq:res1}
\end{equation}
where the term (**) represents the deterministic component of the residuals. The matrix $\mathbf{A}_{n}(\mathbf{I}_{n}-\mathbf{A}_n)^{-1}$ has nonzero entries only in its upper-left $m \times m$ submatrix, due to the structure of $\mathbf{A}_n$. 
Let 
$g_{ij}$, for $i,j = 1, ..., m$, denote the $(i,j)$th entry of 
$\mathbf{A}_n(\mathbf{I}_{n}-\mathbf{A}_n)^{-1}$. 
Depending on the structure of the adjacency matrix $\mathbf{A}_n$, the row sums of $g_{ij}$ diverge or remain bounded as $n$ increases.

In the paper, we use the potential peers' exogenous characteristics, $x_{jt}$, as IVs for their outcomes, $y_{jt}$. 
The vector $\mathbf{Z}_{it}$ collects the characteristics of individual $i$'s potential peers such that $\mathbf{Z}_{it}=[x_{jt}]_{j\in \mathcal{N}_{i}}=(x_{1t},\cdots
,x_{i-1,t},x_{i+1,t},\cdots ,x_{nt})$ and the IV matrix for period $t$, denoted by $\mathbf{Q}_t$, is defined as the set of linearly independent columns in $[\mathbf{Z}_t,\mathbf{x}_t]$, where $\mathbf{Z}_{t}=(\mathbf{e}_{1}\mathbf{Z}_{1t},\cdots
,\mathbf{e}_{n}\mathbf{Z}_{nt})$ and $\mathbf{e}_{i}$ denotes the $i$th column of the identity matrix $\mathbf{I}_{n}$. 

To simplify our analysis here, we replace $\mathbf{x}_t$ in the IV matrix with the matrix  $(\mathbf{e}_{1}x_{1t},$ $\cdots,$ $\mathbf{e}_{n}x_{nt})$. Since the original $\mathbf{x}_t$ can be expressed as a linear combination of the columns of this matrix, the informational content of the IV set remains unchanged. With the modification, $\mathbf{Q}_t =\mathbf{I}_n \otimes \mathbf{x}_t^{\prime}$, which greatly simplifies our analysis, providing clear insight into the power of our test.

Let $\widetilde{\mathbf{u}} = \left(\widetilde{\mathbf{u}}^{\prime}_1, \cdots, \widetilde{\mathbf{u}}^{\prime}_T\right)^{\prime}$, and other terms without time subscripts are defined similarly. Then, our AR test statistic under the alternative can be written as
\begin{eqnarray*}
\frac{1}{\sqrt{K\Phi }}\widetilde{\mathbf{u}}^{\prime }(\mathbf{P}-\mathbf{D})\widetilde{\mathbf{u}} &=&\frac{\beta^2}{%
\sqrt{K\Phi }} \mathbf{x}^{\prime }(\mathbf{I}_{nT}-\mathbf{A}^{\prime
})^{-1}\mathbf{A}^{\prime }(\mathbf{P}-\mathbf{D})\mathbf{A}(\mathbf{I}_{nT}-\mathbf{A})^{-1}\mathbf{x}\\
&&+\frac{2\beta}{\sqrt{K\Phi}} \mathbf{x}^{\prime }(\mathbf{I}_{nT}-\mathbf{A}^{\prime
})^{-1}\mathbf{A}^{\prime}(\mathbf{P}-\mathbf{D})(\mathbf{I}_{nT}-\mathbf{A})^{-1}\mathbf{u} \\
&&+\frac{1}{\sqrt{K\Phi }}\mathbf{u}^{\prime}(\mathbf{I}_{nT}-\mathbf{A}^{\prime
})^{-1}(\mathbf{P}-\mathbf{D})(\mathbf{I}_{nT}-\mathbf{A})^{-1}\mathbf{u}\\
& \equiv & \Xi_1 + \Xi_2 + \Xi_3,
\end{eqnarray*}%
where $\mathbf{A} = \mathbf{I}_T \otimes\mathbf{A}_n$, $\mathbf{P} = \mathbf{Q}(\mathbf{Q}^{\prime}\mathbf{Q})^{-1}\mathbf{Q}^{\prime}$ and $\mathbf{D}$ is a diagonal matrix containing the diagonal
elements of $\mathbf{P}$. 

The first term, $\Xi_1$, is deterministic and contains the squared elements of the vector (**) in Equation (\ref{eq:res1}). The other terms are stochastic with a zero mean.\footnote{It follows directly from the exogeneity of $\mathbf{x}$ that $\text{E}(\Xi_2) = 0$. As $\mathbf{Q}_t = \mathbf{I}_n \otimes \mathbf{x}_{t}^{\prime}$ and $\mathbf{Q} = [\mathbf{Q}^{\prime}_1, \cdots, \mathbf{Q}^{\prime}_T ]^{\prime}$, the projection matrix 
$\mathbf{P} = \mathbf{Q}(\mathbf{Q}^{\prime}\mathbf{Q})^{-1}\mathbf{Q}^{\prime} = [\mathbf{X}^{\prime}(\mathbf{X}\mathbf{X}^{\prime})^{-1}\mathbf{X}]\otimes  \mathbf{I}_n$ where $\mathbf{X}=[\mathbf{x}_{1},\cdots,\mathbf{x}_{T}]$. Consequently, $\mathbf{P}-\mathbf{D}$ is a block matrix with zero diagonal blocks. Furthermore, since $(\mathbf{I}_{nT}-\mathbf{A})^{-1}$ is a block diagonal matrix and $u_{it}$ are independent across $t$, $\text{E}(\Xi_3) = 0$.} Therefore, for the AR test to be consistent, the deterministic component $\Xi_1$ must diverge. In the following, we derive the conditions under which the deterministic component diverges to infinity. The sufficient conditions for the stochastic terms to be bounded are relegated to the online appendix for brevity. 

Since  $\mathbf{Q}_t = \mathbf{I}_n \otimes \mathbf{x}_t^{\prime}$ is a collection of all basis vectors for individual characteristics in period $t$, $\mathbf{A}_n(\mathbf{I}_{n}-\mathbf{A}_n)^{-1}\mathbf{x}_t$ can be viewed as a linear combination of the columns of $\mathbf{Q}_t$.\footnote{This can be seen from $\mathbf{A}_n(\mathbf{I}_{n}-\mathbf{A}_n)^{-1}\mathbf{x}_t 
= \mathbf{Q}_t \text{vec}\left([\mathbf{A}_n(\mathbf{I}_{n}-\mathbf{A}_n)^{-1}]^{\prime}\right)$, where 
vec($\cdot$) denotes the vectorization of a matrix by stacking its columns on top of one another.} Consequently,
\begin{equation*}
    \mathbf{x}^{\prime }(\mathbf{I}_{nT}-\mathbf{A}^{\prime})^{-1}\mathbf{A}^{\prime }\mathbf{P}\mathbf{A}(\mathbf{I}_{nT}-\mathbf{A})^{-1}\mathbf{x}
    =\mathbf{x}^{\prime }(\mathbf{I}_{nT}-\mathbf{A}^{\prime})^{-1}\mathbf{A}^{\prime }\mathbf{A}(\mathbf{I}_{nT}-\mathbf{A})^{-1}\mathbf{x}.
\end{equation*}
On the other hand, we have
\begin{equation*}
    \mathbf{x}^{\prime }(\mathbf{I}_{nT}-\mathbf{A}^{\prime})^{-1}\mathbf{A}^{\prime }\mathbf{D}\mathbf{A}(\mathbf{I}_{nT}-\mathbf{A})^{-1}\mathbf{x}
    \leq C_p\mathbf{x}^{\prime }(\mathbf{I}_{nT}-\mathbf{A}^{\prime})^{-1}\mathbf{A}^{\prime }\mathbf{A}(\mathbf{I}_{nT}-\mathbf{A})^{-1}\mathbf{x},
\end{equation*}
where $C_p < 1$ is the upper bound of the diagonal elements of $\mathbf{P}$ defined in Assumption 2 of the paper.  
Hence, 
\begin{eqnarray*}
\Xi_1 
&\geq& \frac{\beta^2 (1 - C_p)}{\sqrt{K \Phi}} 
    \mathbf{x}^{\prime} (\mathbf{I}_{nT} - \mathbf{A}^{\prime})^{-1} 
    \mathbf{A}^{\prime}  \mathbf{A} 
    (\mathbf{I}_{nT} - \mathbf{A})^{-1} \mathbf{x}\\
&=&  \frac{\beta^2 (1 - C_p)}{\sqrt{K \Phi}} 
    \sum_{t=1}^{T} \sum_{i=1}^{m} 
    \left( \sum_{j=1}^{m} g_{ij} x_{jt} \right)^2.
\end{eqnarray*}

Therefore, the consistency of our AR test requires (\ref{con1}) in the main text to hold.
The term in (\ref{con1}) of the main text
\begin{eqnarray*} \label{cp}
\beta^2 \sum_{t=1}^{T} \sum_{i=1}^{m}\left(\sum_{j=1}^{m} g_{ij}x_{jt} \right)^2
\end{eqnarray*}
is equivalent to summing the squared elements of the vector (**) in Equation (\ref{eq:res1}) over individuals and time periods. It quantifies the strength of peer effects in the data and thus determines the power of our test. Accordingly, we refer to (\ref{con1}) in the main text as the power formula. Given $K = O(n^2)$ in this setup and under the maintained assumption $x_{it} = O_p(1)$, in Remark \ref{power}, we discuss the power of our test for several representative network examples using the power formula.

\begin{remark*} \label{beta-estimation}
In the preceding power analysis, we impose a simplifying assumption that $\beta$ is known. When $\beta$ is instead estimated via ordinary least squares, the residual vector $\widetilde{\mathbf{u}}$ is given by
\begin{equation}
\widetilde{\mathbf{u}}=\mathbf{M}_x\mathbf{y}
 =   \underbrace{\beta\mathbf{M}_x\mathbf{A}(\mathbf{I}_{nT} - \mathbf{A})^{-1}\mathbf{x}}_{\#\#} + \mathbf{M}_x(\mathbf{I}_{nT} - \mathbf{A})^{-1}\mathbf{u}, \label{eq:res2}
\end{equation}
where $\mathbf{M}_x = \mathbf{I}_{nT} - \mathbf{x} \left(\mathbf{x}^{\prime}\mathbf{x} \right)^{-1}\mathbf{x}^{\prime}$.

The preceding power analysis shows that when $\beta$ is known, the test’s power comes from the total variation of the (**) term in Equation (\ref{eq:res1}), which captures the overall peer effects in the data.
When $\beta$ is estimated, the (**) term in Equation (\ref{eq:res1}) becomes the (\#\#) term in Equation (\ref{eq:res2}), representing the projection residual of the (**) term onto $\mathbf{x}$.
Therefore, when $\beta$ is estimated, the power of the test may be lower if some variation of the (**) term (i.e., variation in peer effects) is explained by $\mathbf{x}$ (i.e., own characteristics).

Using similar algebra as above, it can be shown that when $\beta$ is estimated, the deterministic component $\Xi_1$ of the AR test statistic is bounded below by
\begin{equation*}
\Xi_1 \geq \frac{\beta^2 (1 - C_p)}{\sqrt{K \Phi}} 
    \mathbf{x}^{\prime} (\mathbf{I}_{nT} - \mathbf{A}^{\prime})^{-1} 
    \mathbf{A}^{\prime} \mathbf{M}_x \mathbf{A} 
    (\mathbf{I}_{nT} - \mathbf{A})^{-1} \mathbf{x}.
\end{equation*}
Therefore, the main conclusions in Remark \ref{power}, particularly those regarding the relative divergence rates of $n$ and $T$, remain unchanged as long as
\[
\frac{\mathbf{x}^{\prime}(\mathbf{I}_{nT} - \mathbf{A}^{\prime})^{-1}\mathbf{A}^{\prime}\mathbf{M}_x\mathbf{A}(\mathbf{I}_{nT} - \mathbf{A})^{-1}\mathbf{x}}
{\mathbf{x}^{\prime}(\mathbf{I}_{nT} - \mathbf{A}^{\prime})^{-1}\mathbf{A}^{\prime}\mathbf{A}(\mathbf{I}_{nT} - \mathbf{A})^{-1}\mathbf{x}},
\]
does not converge to zero; or, more intuitively, the $R^2$ from regressing $\mathbf{A}(\mathbf{I}_{nT} - \mathbf{A})^{-1}\mathbf{x}$ on $\mathbf{x}$ does not converge to one, as the sample size increases.
Recall that the $i$th element of $\mathbf{A}(\mathbf{I}_{nT} - \mathbf{A})^{-1}\mathbf{x}$ is $\sum_{t=1}^{T} \sum_{j=1}^{m} g_{ij} x_{jt}$. Hence, $\mathbf{A}(\mathbf{I}_{nT} - \mathbf{A})^{-1}\mathbf{x}$ is generally not perfectly collinear with $\mathbf{x}$ under the alternative because (i) $g_{ii}x_{it}$ cannot be perfectly predicted by $x_{it}$ when $g_{ii}$ varies across individuals, and (ii) the other terms, $g_{ij}x_{jt}$, cannot be perfectly explained by $x_{it}$ when $x_{it}$ varies sufficiently across individuals.
$\blacksquare$

\end{remark*}

\section{Estimation of Excess Kurtosis of Regression Error} \label{exk}
We propose the following estimator for the excess kurtosis of $\epsilon_{it}$, i.e., $\mu_4 - 3 \sigma^4$:
\begin{equation*}
\widehat{\kappa} = \widehat{\mu}_4 - 3 \widehat{\sigma}^4 \left(\frac{\pi_{1}}{\pi_{2}} \right),
\end{equation*}
where
\begin{eqnarray*}
&& \widehat{\mu}_4 = \sum_{i=1}^n\sum_{t=1}^T \widehat{\epsilon}^{*4}_{it}/\pi_2, \quad \widehat{\sigma}^2 = \sum_{i=1}^n\sum_{t=1}^T \widehat{\epsilon}^{*2}_{it}/N^{*}, \quad N^{*} = (n-1)(T-1), \quad N = nT,\\
&& \pi_{1} = \sum_{r=1}^N (J_{rr})^2 = N^{*2}/N, \quad \pi_{2} = \sum_{r=1}^N \sum_{s=1}^N (J_{rs})^4 =  N^{*}(N^{*3} + (n-1)^3 + (T-1)^3 + 1)/N^3,\\
& & \text{$J_{rs}$ is the $(r,s)$th element of} \,\, \mathbf{J} =(\mathbf{I}_{T}-T^{-1}\boldsymbol{\iota} _{T}\boldsymbol{\iota} _{T}^{\prime })\otimes (\mathbf{I}_{n}-n^{-1}\boldsymbol{\iota} _{n}\boldsymbol{\iota} _{n}^{\prime }),\\
&& \widehat{\epsilon}^{*}_{it} = \epsilon_{it}^{*} + \mathbf{X}^{*}_{it}(\boldsymbol{\beta} - \widehat{\boldsymbol{\beta}}), \quad \epsilon_{it}^{*} =  \epsilon_{it} - \bar{ \epsilon}_{i.} - \bar{ \epsilon}_{.t} +\bar{ \epsilon}_{..}, \,\, \bar{ \epsilon}_{i.} = \sum_{t=1}^{T} \epsilon_{it}/T, \,\, \bar{ \epsilon}_{.t} = \sum_{i=1}^{n} \epsilon_{it}/n,\\
&& \bar{ \epsilon}_{..} = \sum_{i=1}^{n} \sum_{t=1}^{T} \epsilon_{it}/N, \, \text{and $\mathbf{X}^{*}_{it}$ is similarly defined.}
\end{eqnarray*}
All asymptotic results in this section hold for $T \to \infty$ and/or $n \to  \infty$. We assume the following conditions for $\mathbf{X}_{it}$:
\begin{assumption} \label{as1}
$\mathbf{X}_{it}$ are i.i.d. over $i$ and $t$, and has finite moments up to order 8.
\end{assumption}
The i.i.d. assumption is to simplify the proof and can be relaxed with more lengthy arguments. See \citet{SW2008} for an example.

\begin{description}
\item[Proposition C.1.]  \label{prop_ek} Suppose Assumptions 1' (in the main text) and \ref{as1} hold. Under $H_{0}:\boldsymbol{\alpha} =0$, $\widehat{\kappa} = \mu_4 - 3 \sigma^4 + o_p(1)$.
\end{description}

\begin{proof}[\textbf{Proof of Proposition C.1.}]
To simply the calculations and notations, we consider the simple case with scalar $x_{it}$, and convert the two-dimensional index $(i,t) = (1,1) ..., (n,T)$ to a one-dimensional index $i = 1, ..., N (= nT)$, while keeping the original ordering. Throughout the proof, we assume $n \geq 2$ and $T \geq 2$, and use the following properties of $\mathbf{J}$ and Lemmas:
\begin{enumerate}
\item [(P1)] For all $i,j$ and $q \geq 1$, $J_{ii} = N^{*}/N$, $|J_{ij}| < 1$, and $\sum_{j=1}^{N}J_{ij} = 0$.
\item [(P2)] Since $\mathbf{J}$ is symmetric \& idempotent, $\sum_{r=1}^N J_{ir} J_{jr} = J_{ij}$, which implies $\sum_{r=1}^N J_{ir}^2 = J_{ii}$.
\end{enumerate}

\begin{lemma} \label{lem1} The elements of $\mathbf{J}$ satisfy the following:
\begin{enumerate} 
\item[(1)] For $q \geq 1$, $0 \leq \sum_{j=1}^{N} J_{ij}^{q} \leq \sum_{j=1}^{N} J_{ij}^2 = N^{*}/N$, and $\sum_{j=1}^{N} |J_{ij}|^{q} \leq \sum_{j=1}^{N} |J_{ij}| = 4N^{*}/N$
\item[(2)]  For $q_1 \geq 1, q_2 \geq 1$, $|\sum_{r=1}^N J_{ir}^{q_1} J_{jr}^{q_2}| \leq |J_{ij}|$.
\end{enumerate}
\end{lemma} 

\begin{lemma} \label{lem2} Suppose Assumption 1' (in the main text) holds. Then, $\mathrm{E}\left[ \left( \sum_{i=1}^{N} \epsilon_i^{*4}/\pi_2\right)^2 \right] = \left( \mu_4 - 3\sigma^4 + 3\sigma^4\pi_1/\pi_2 \right)^2 + o(1).$
\end{lemma} 
The proofs of the Lemmas are in the supplementary material.

To prove the consistency of $\widehat{\kappa}$, we first show $\widehat{\kappa} = \widetilde{\kappa} +o_{p}(1),$ where $\widetilde{\kappa}$ is defined as $\widehat{\kappa}$ with  $\widehat{\epsilon}^{*}_{i}$ replaced by $\epsilon_{i}^{*}$. The statistics $\widetilde{\mu}$ and $\widetilde{\sigma}^4$ are similarly defined. First note that
\begin{eqnarray*}
\widehat{\mu}_4 &=& \frac{N}{\pi_{2}}\sum_{i=1}^N \widehat{\epsilon}^{*4}_{i}/N \nonumber \\
&=&  \frac{N}{\pi_{2}} \left[\sum_{i=1}^N \epsilon_{i}^{*4}/N + 4\left(\sum_{i=1}^N\epsilon_{i}^{*3}x_{i}^{*}/N\right)(\boldsymbol{\beta}-\widehat{\boldsymbol{\beta}}) + 6\left(\sum_{i=1}^N\epsilon_{i}^{*2}x_{i}^{*2}/N\right)(\boldsymbol{\beta}-\widehat{\boldsymbol{\beta}})^2 \right. \nonumber\\
&& \left. + 4\left(\sum_{i=1}^N\epsilon_{i}^{*}x_{i}^{*3}/N\right)(\boldsymbol{\beta}-\widehat{\boldsymbol{\beta}})^3 + \left(\sum_{i=1}^N x_{i}^{*4}/N\right)(\boldsymbol{\beta}-\widehat{\boldsymbol{\beta}})^4 \right].
\end{eqnarray*}
Since $\boldsymbol{\beta}-\widehat{\boldsymbol{\beta}} = O(1/\sqrt{N})$, it suffices to show that the four terms, $\sum_{i=1}^N\epsilon_{i}^{*3}x_{i}^{*}/N,$ $\sum_{i=1}^N\epsilon_{i}^{*2}x_{i}^{*2}/N,$ $\sum_{i=1}^N\epsilon_{i}^{*}x_{i}^{*3}/N$ and $\sum_{i=1}^N x_{i}^{*4}/N$ (hereafter, ``the four terms") are $O_p(1)$ to show $\widehat{\kappa} = \widetilde{\kappa} +o_{p}(1)$.

The transformed error $\epsilon^{*}_{i}$ is the $i^{th}$ element of $\mathbf{J}\boldsymbol{\epsilon}$, where $\boldsymbol{\epsilon}$ is the error vector, so it can be written as $\epsilon^{*}_{i} = \sum_{j=1}^{N}J_{ij}\epsilon_{j}$. Then,
\begin{eqnarray}
\mathrm{E}\left(\sum_{i=1}^N \epsilon_{i}^{*4}/N \right)& = & \sum_{i=1}^N  \mathrm{E}\left(\sum_{j=1}^{N}\sum_{k=1}^{N}\sum_{l=1}^{N}\sum_{m=1}^{N} J_{ij} J_{ik} J_{il} J_{im}u_{j} u_{k} u_{l} u_{m}\right)/N \nonumber \\
& =&  \mu_4 \sum_{i=1}^{N} \sum_{j=1}^N J_{ij}^4/N + 3\sigma^4 \sum_{i=1}^N \sum_{j=1}^{N} \sum_{k \ne j} J_{ij}^2 J_{ik}^2/N \nonumber \\
&= & (\mu_4 - 3\sigma^4) \pi_2/N + 3\sigma^4 \sum_{i=1}^N \sum_{j=1}^{N} \sum_{k=1}^{N} J_{ij}^2 J_{ik}^2/N  \nonumber\\
& =& (\mu_4 - 3\sigma^4) \pi_2/N + 3\sigma^4 \pi_1/N, \label{eq2}
\end{eqnarray}
which is bounded since $\epsilon^{}_{it}$ has finite moments up to order 8, and $\pi_1/N = O(1)$ and $\pi_2/N = O(1)$. The second and last equalities are due to that $\epsilon_{it}$ are $i.i.d.$ and P2. 

Note that the result (\ref{eq2}) implies $\mathrm{E}(\epsilon_{i}^{*4})$ is finite and constant over $i$. Also, it can be easily seen from the proof of Lemma \ref{lem2} that $\mathrm{E}(\epsilon_{i}^{*8})$ is finite and constant over $i$. Similarly, it can be shown that both $\epsilon_{i}^{*}$ and $x_{i}^{*}$ have finite moments up to order 8, which implies that the expectations of the four terms are bounded. For example, $\mathrm{E}\left(\sum_{i=1}^N\epsilon_{i}^{*3}x_{i}^{*}/N\right) = \sum_{i=1}^N \mathrm{E}(\epsilon_{i}^{*3}x_{i}^{*})/N$, which is bounded since $|\mathrm{E}(\epsilon_{i}^{*3}x_{i}^{*})| \leq \mathrm{E}(|\epsilon_{i}^{*3}x_{i}^{*}|) \leq \sqrt{\mathrm{E}(\epsilon_{i}^{*6})\mathrm{E}(x_{i}^{*2})}$ by the Holder inequality. The same argument can be applied to the other terms.

Also, the variances of the four terms are bounded. For example, $\mathrm{Var}\left(\sum_{i=1}^N\epsilon_{i}^{*3}x_{i}^{*}/N\right) \leq (\sum_{i=1}^N \sqrt{\mathrm{Var} \left(\epsilon_{i}^{*3}x_{i}^{*} \right)}/N)^2$ by the covariance inequality, which is bounded since $\mathrm{Var}(\epsilon_{i}^{*3}x_{i}^{*}) \leq \mathrm{E}(\epsilon_{i}^{*6}x_{i}^{*2}) \leq \mathrm{E}(\epsilon_{i}^{*8})^{6/8} \mathrm{E}(x_{i}^{*8})^{1/4}$ by the Holder inequality.

Then, it follows that $\widehat{\mu} = \widetilde{\mu} +o_{p}(1)$, and the same result can be obtained for $\widehat{\sigma}^4$ by applying the same arguments, from which the desired result follows.\\

Next, we show $\widetilde{\kappa}  = \mu_4 - 3 \sigma^4 + o_p(1)$ by showing that (i) $\mathrm{E}(\widetilde{\kappa}) =  \mu_4 - 3\sigma^4 + o(1)$ and (ii) $\mathrm{Var}(\widetilde{\kappa}) = o(1)$. 

(i) Similarly to (\ref{eq2}),
\begin{eqnarray}
 \mathrm{E}\left[ \left(\sum_{i=1}^{N} \epsilon_i^{*2}/N^{*} \right)^2\right]& = &  \mathrm{E}\left[ \left( \boldsymbol{\epsilon}^{\prime} \mathbf{J} \boldsymbol{\epsilon} \right)^2\right]/N^{*2} \nonumber \\
& = & (\mu_4 - 3\sigma^4) \pi_{1}/N^{*2}  + \sigma^4 (1 + 2/N^{*} ). \label{eq3}
\end{eqnarray}
Then, the results (\ref{eq2}) and (\ref{eq3}) yield 
\begin{eqnarray*}
 \mathrm{E}(\widetilde{\kappa})& = & \frac{ \mathrm{E}\left(\sum_{i=1}^{N} \epsilon_i^{*4}\right)}{\pi_{2}} - 3 \frac{ \mathrm{E}\left[ \left(\sum_{i=1}^{N} \epsilon_i^{*2} \right)^2\right]}{N^{*2}} \left(\frac{\pi_{1}}{\pi_{2}} \right)  \\
& = & \mu_4 - 3\sigma^4  - 3\left(\frac{\pi_{1}}{\pi_{2}} \right) \left[ (\mu_4 - 3\sigma^4)\frac{\pi_{1}}{N^{*2}} + 2\frac{\sigma^4}{N^{*}} \right]\\
& = & \mu_4 - 3\sigma^4 + o(1),
\end{eqnarray*}
where the last equality is due to $\pi_{1}/N^{*} = O(1)$ and $\pi_{1}/\pi_{2} = O(1)$.

(ii) Since $ \mathrm{Var}(a + b) \leq (\sqrt{ \mathrm{Var}(a)} + \sqrt{ \mathrm{Var}(b)})^2$, it suffices to show $ \mathrm{Var}(\widetilde{\mu}) = o(1)$ and $ \mathrm{Var}(\widetilde{\sigma}^4) = o(1)$ to show $ \mathrm{Var}(\widetilde{\kappa}) = o(1)$. First, Lemma \ref{lem2} and result (\ref{eq2}) yield that $ \mathrm{Var}(\widetilde{\mu}) = o(1)$. Also, the result (\ref{eq3}) and that $ \mathrm{E}(\widetilde{\sigma}^2) = \sigma^2$ imply $\widetilde{\sigma}^2 = \sigma^2 + o_p(1)$. Then, by the Slutsky's theorem, we have $\widetilde{\sigma}^4 = \sigma^4 + o_p(1)$, which implies $ \mathrm{Var}(\widetilde{\sigma}^4) = o(1)$. This completes the proof.\end{proof}

\numberwithin{table}{section}

\section{Additional Simulations: Power of $\texttt{T}_{\texttt{JL}}$ and $\texttt{T}_{\texttt{AG}}$} \label{addsimul} 

\begin{table}[H] \singlespacing
\centering
\begin{threeparttable}
\caption{Power of $\texttt{T}_{\texttt{JL}}$ and $\texttt{T}_{\texttt{AG}}$} \label{power_JL_AG}
\begin{tabular}{cc|cc|cc} \hline \hline
      &      & \multicolumn{2}{c}{Normal} & \multicolumn{2}{|c}{Log-Normal} \\
$\rho$   & $\beta$ & $\texttt{T}_{\texttt{JL}}$           & $\texttt{T}_{\texttt{AG}}$          & $\texttt{T}_{\texttt{JL}}$            & $\texttt{T}_{\texttt{AG}}$            \\ \hline
0.00     & 1.0    & 0.038        & 0.038       & 0.042         & 0.095         \\
0.05  & 1.0    & 0.078        & 0.078       & 0.075         & 0.148         \\
0.10   & 1.0    & 0.468        & 0.470       & 0.470         & 0.488         \\
0.30   & 1.0    & 0.996        & 0.996       & 0.981         & 0.985         \\
0.50   & 1.0    & 0.991        & 0.992       & 0.985         & 0.986         \\
-0.05 & 1.0    & 0.078        & 0.077       & 0.068         & 0.140         \\
-0.10  & 1.0    & 0.318        & 0.318       & 0.285         & 0.397         \\
-0.30  & 1.0    & 0.998        & 0.998       & 0.990         & 0.991         \\
-0.50  & 1.0    & 0.990        & 0.990       & 0.986         & 0.986         \\ \hline
0.30   & 0.0    & 0.169        & 0.169       & 0.164         & 0.193         \\
0.30   & 0.1  & 0.183        & 0.184       & 0.181         & 0.210         \\
0.30   & 0.3  & 0.295        & 0.296       & 0.278         & 0.314         \\
0.30   & 0.5  & 0.558        & 0.559       & 0.537         & 0.570         \\
0.30   & 0.7  & 0.858        & 0.859       & 0.820         & 0.839         \\
0.30   & 0.9  & 0.982        & 0.982       & 0.963         & 0.970         \\
0.30   & 1.0    & 0.995        & 0.995       & 0.980         & 0.985         \\
0.30   & -0.1 & 0.183        & 0.184       & 0.172         & 0.199         \\
0.30   & -0.3 & 0.303        & 0.304       & 0.295         & 0.325         \\
0.30   & -0.5 & 0.563        & 0.563       & 0.557         & 0.591         \\
0.30   & -0.7 & 0.860        & 0.861       & 0.831         & 0.851         \\
0.30   & -0.9 & 0.981        & 0.981       & 0.953         & 0.962         \\
0.30   & -1.0   & 0.996        & 0.996       & 0.981         & 0.986        
  \\ \hline \hline
\end{tabular}
\begin{tablenotes}
 \small
 \item Note: $(n,T) = (30,50),$ $\texttt{ND} = 0.3$ and $5\%$ significance level test. All tests use the same set of IVs and two-way within-transformed residuals described in Sections \ref{ARwof} and \ref{ARwf}. Chi-square distribution to calculate the critical values.
  \end{tablenotes}
  \end{threeparttable}
\end{table}

$\texttt{T}_{\texttt{JL}}$ is the proposed AR test and $\texttt{T}_{\texttt{AG}}$ is the \citet{AG2011}'s J test, which allows the number of IVs to grow as fast as $\texttt{T}_{\texttt{JL}}$, but assumes the balanced covariate design. In Table \ref{power_JL_AG}, the two tests exhibit almost the same rejection rates in many settings, particularly under normal errors, indicating that there is little or no power loss when using $\texttt{T}_{\texttt{JL}}$, instead of $\texttt{T}_{\texttt{AG}}$, for testing the presence of peer effects without the balanced covariate design assumption.
 
\bibliographystyle{agsm}
\bibliography{\jobname}      

@article{AS2019,
 author = {Anatolyev, Stanislav},
 journal = {Journal of Economic Surveys},
 number = {2},
 pages = {689-726},
 title = {MANY INSTRUMENTS AND/OR REGRESSORS: A FRIENDLY GUIDE},
 volume = {33},
 year = {2019}}

@article{ANA2012,
title = {Inference in regression models with many regressors},
journal = {Journal of Econometrics},
volume = {170},
number = {2},
pages = {368-382},
year = {2012},
author = {Stanislav Anatolyev}
}

@article{CAL2011,
title = {Hypothesis testing in linear regression when k/n is large},
journal = {Journal of Econometrics},
volume = {165},
number = {2},
pages = {163-174},
year = {2011},
author = {Gray Calhoun}
}

@article{Catt2018,
author = {Matias D. Cattaneo and Michael Jansson and Whitney K. Newey},
title = {Inference in Linear Regression Models with Many Covariates and Heteroscedasticity},
journal = {Journal of the American Statistical Association},
volume = {113},
number = {523},
pages = {1350--1361},
year = {2018}
}

@article{Catt2018a,
author = {Matias D. Cattaneo and Michael Jansson and Xinwei Ma},
title = {Two-Step Estimation and Inference with Possibly Many Included Covariates},
journal = {The Review of Economic Studies},
volume = {86},
issue = {3},
pages = {1095--1122},
year = {2018}
}

@article{PY2021,
title = {Estimation and inference in spatial models with dominant units},
journal = {Journal of Econometrics},
volume = {221},
number = {2},
pages = {591-615},
year = {2021},
author = {M. Hashem Pesaran and Cynthia Fan Yang}
}

@article{GO2012,
author = {Gibbons, Stephen and Overman, Henry G.},
title = {MOSTLY POINTLESS SPATIAL ECONOMETRICS?},
journal = {Journal of Regional Science},
volume = {52},
number = {2},
pages = {172-191},
year = {2012}
}

@article{AG2011,
 author = {Anatolyev, Stanislav and Gospodinov, Nikolay},
 journal = {Econometric Theory},
 number = {2},
 pages = {427--441},
 title = {SPECIFICATION TESTING IN MODELS WITH MANY INSTRUMENTS},
 volume = {27},
 year = {2011}}

@article{AS2023,
title = {Testing many restrictions under heteroskedasticity},
journal = {Journal of Econometrics},
volume = {236},
number = {1},
pages = {105473},
year = {2023},
author = {Stanislav Anatolyev and Mikkel Sølvsten},
}

@article{AY2017,
title={ASYMPTOTICS OF DIAGONAL ELEMENTS OF PROJECTION MATRICES UNDER MANY INSTRUMENTS/REGRESSORS},
volume={33},
number={3},
journal={Econometric Theory},
author={Anatolyev, Stanislav and Yaskov, Pavel},
year={2017},
pages={717–738}}

@article{Angrist1999,
 author = {J. D. Angrist and G. W. Imbens and A. B. Krueger},
 journal = {Journal of Applied Econometrics},
 number = {1},
 pages = {57--67},
 title = {Jackknife Instrumental Variables Estimation},
 volume = {14},
 year = {1999}}

@article{BFK2012,
 author = {Badi H. Baltagi and Qu Feng and Chihwa Kao},
 journal = {Journal of Econometrics},
 number = {1},
 pages = {164-177},
 title = {A \uppercase{L}agrange \uppercase{M}ultiplier test for cross-sectional dependence in a fixed effects panel data model},
 volume = {170},
 year = {2012},}

@article{BPR2021,
 author = {Battaglini, Marco and Patacchini, Eleonora and Rainone, Edoardo},
 journal = {The Review of Economic Studies},
 month = {09},
 number = {4},
 pages = {1694-1747},
 title = {Endogenous Social Interactions with Unobserved Networks},
 volume = {89},
 year = {2021}}

@article{Bekker1994,
 author = {Bekker, Paul A.},
 journal = {Econometrica},
 number = {3},
 pages = {657--681},
 title = {Alternative Approximations to the Distributions of Instrumental Variable Estimators},
 volume = {62},
 year = {1994}}

@article{Bekker2015,
title = {Jackknife instrumental variable estimation with heteroskedasticity},
journal = {Journal of Econometrics},
volume = {185},
number = {2},
pages = {332-342},
year = {2015},
author = {Paul A. Bekker and Federico Crudu}
}

@article{Berri99,
 author = {Berri, D. J.},
 journal = {Managerial and Decision Economics},
 pages = {411-427},
 title = {Who is Most Valuable? Measuring the Player's Production of Wins in the \uppercase{NBA}},
 volume = {20},
 number = {8},
 year = {1999}}

@article{Blume2015,
 author = {Blume, Lawrence E. and Brock, William A. and Durlauf, Steven N. and Jayaraman, Rajshri},
 journal = {Journal of Political Economy},
 number = {2},
 pages = {444-496},
 title = {Linear Social Interactions Models},
 volume = {123},
 year = {2015}}

@unpublished{Bonaldi2015,
 author = {Bonaldi, Pietro and Horta{\c c}su, Ali and Kastl, Jakub},
 title = {An Empirical Analysis of Funding Costs Spillovers in the \uppercase{EURO}-zone with Application to Systemic Risk},
 note = {Working paper (http://www.nber.org/papers/w21462)},
 year = {2015}}

@article{BDF09,
 author = {Bramoull\'{e}, Y. and H. Djebbari and B. Fortin},
 journal = {Journal of Econometrics},
 pages = {41-55},
 title = {Identification of peer effects through social networks},
 volume = {150},
        number = {1},
 year = {2009}}

@article{Breza2020,
Author = {Breza, Emily and Chandrasekhar, Arun G. and McCormick, Tyler H. and Pan, Mengjie},
Title = {Using Aggregated Relational Data to Feasibly Identify Network Structure without Network Data},
Journal = {American Economic Review},
Volume = {110},
Number = {8},
Year = {2020},
Pages = {2454-84}}

@article{Chao2014,
 author = {John C. Chao and Jerry A. Hausman and Whitney K. Newey and Norman R. Swanson and Tiemen Woutersen},
 journal = {Journal of Econometrics},
 pages = {15-21},
 title = {Testing overidentifying restrictions with many instruments and heteroskedasticity},
 volume = {178},
 year = {2014}}

@article{Chao2012,
 author = {Chao, John C. and Swanson, Norman R. and Hausman, Jerry A. and Newey, Whitney K. and Woutersen, Tlemen},
 journal = {Econometric Theory},
 number = {1},
 pages = {42--86},
 title = {ASYMPTOTIC DISTRIBUTION OF \uppercase{JIVE} IN A HETEROSKEDASTIC \uppercase{IV} REGRESSION WITH MANY INSTRUMENTS},
 volume = {28},
 year = {2012}}

@article{CGL2012,
 author = {Chen, Jia and Gao, Jiti and Li, Degui},
 journal = {Econometric Theory},
 number = {5},
 pages = {1144--1163},
 title = {A NEW DIAGNOSTIC TEST FOR CROSS-SECTION UNCORRELATEDNESS IN NONPARAMETRIC PANEL DATA MODELS},
 volume = {28},
 year = {2012}}

@article{Crudu2021,
 author = {Crudu, Federico and Mellace, Giovanni and S{\'a}ndor, Zsolt},
 journal = {Econometric Theory},
 number = {2},
 pages = {281-310},
 title = {INFERENCE IN INSTRUMENTAL VARIABLE MODELS WITH HETEROSKEDASTICITY AND MANY INSTRUMENTS},
 volume = {37},
 year = {2021}}

@article{DRS24,
    author = {de Paula, {\'A}ureo and Rasul, Imran and Souza, Pedro C L},
    title = {Identifying Network Ties from Panel Data: Theory and an Application to Tax Competition},
    journal = {The Review of Economic Studies},
    pages = {2691-2729},
 volume = {92},
    number = {4},
    year = {2024}
}

@article{Donald2003,
 author = {Stephen G. Donald and Guido W. Imbens and Whitney K. Newey},
 journal = {Journal of Econometrics},
 number = {1},
 pages = {55-93},
 title = {Empirical likelihood estimation and consistent tests with conditional moment restrictions},
 volume = {117},
 year = {2003}}

@article{er1959,
  title={On random graphs \uppercase{I}},
  author={Erd{\"{o}}s, Paul and R{\'e}nyi, Alfr{\'e}d},
  journal={Publicationes Mathematicae Debrecen},
  volume={6},
  pages={290-297},
  year={1959}
}

@article{EK2007,
 author = {Ertur, Cem and Koch, Wilfried},
 journal = {Journal of Applied Econometrics},
 number = {6},
 pages = {1033-1062},
 title = {Growth, technological interdependence and spatial externalities: theory and evidence},
 volume = {22},
 year = {2007}}

@article{Hansen2008,
 author = {Christian Hansen and Jerry Hausman and Whitney Newey},
 journal = {Journal of Business \& Economic Statistics},
 number = {4},
 pages = {398--422},
 title = {Estimation with Many Instrumental Variables},
 volume = {26},
 year = {2008}}

@article{Hausman2012,
author = {Hausman, Jerry A. and Newey, Whitney K. and Woutersen, Tiemen and Chao, John C. and Swanson, Norman R.},
title = {Instrumental variable estimation with heteroskedasticity and many instruments},
journal = {Quantitative Economics},
volume = {3},
number = {2},
pages = {211-255},
year = {2012}
}

@article{HO2013,
 author = {Chun-Yu Ho and Wei Wang and Jihai Yu},
 journal = {Economics Letters},
 number = {3},
 pages = {450-453},
 title = {Growth spillover through trade: A spatial dynamic panel data approach},
 volume = {120},
 year = {2013}}

@article{HJS20,
 author = {Horrace, W. C. and Jung, H. and Sanders, S},
 journal = {Journal of Business \& Economic Statistics},
 number = {1},
 pages = {35-49},
 title = {Network Competition and Team Chemistry in the \uppercase{NBA}},
 volume = {40},
 year = {2022}}

@article{Lee07group,
 author = {Lee, L.-F.},
 journal = {Journal of Econometrics},
 pages = {333-374},
 title = {Identification and estimation of econometric models with group interactions, contextual factors and fixed effects},
 volume = {140},
 number = {2},
 year = {2007}}

@article{LYY23,
author = {L.-F. Lee and C. Yang and J. Yu},
title = {\uppercase{QML} and Efficient \uppercase{GMM} Estimation of Spatial Autoregressive Models with Dominant (Popular) Units},
journal = {Journal of Business \& Economic Statistics},
volume = {41},
number = {2},
pages = {550-562},
year = {2023},
}

@article{LO2012,
author = {Yoonseok Lee and Ryo Okui},
title = {{Hahn-Hausman} test as a specification test},
journal = {Journal of Econometrics},
volume = {167},
number = {1},
pages = {133-139},
year = {2012}
}

@article{LQT23,
 author = {Lewbel, Arthur and Qu, Xi and Tang, Xun},
 journal = {Journal of Political Economy},
 number = {4},
 pages = {898-946},
 title = {Social Networks with Unobserved Links},
 volume = {131},
 year = {2023}}

@article{LP2018,
 author = {Liu, X. and Prucha, I. R.},
 journal = {Journal of Econometrics},
 number = {1},
 pages = {92-113},
 title = {A Robust Test for Network Generated Dependence},
 volume = {207},
 year = {2018}}

@article{LP2025,
title = {On testing for spatial or social network dependence in panel data allowing for network variability},
journal = {Journal of Econometrics},
volume = {247},
pages = {105925},
year = {2025},
author = {Xiaodong Liu and Ingmar R. Prucha},
}

@unpublished{Manresa2016,
 author = {Elena Manresa},
 title = {Estimating the Structure of Social Interactions Using Panel Data},
 note = {Working paper},
 year = {2016}}

@article{Manski93,
 author = {Manski, C. F.},
 journal = {The Review of Economic Studies},
 pages = {531-542},
 title = {Identification of endogenous social effects: the reflection problem},
 volume = {60},
 number = {3},
 year = {1993}}

@article{MS2022,
 author = {Mikusheva, Anna and Sun, Liyang},
 journal = {The Review of Economic Studies},
 number = {5},
 pages = {2663-2686},
 title = {Inference with Many Weak Instruments},
 volume = {89},
 year = {2022}}

@article{Moran,
 author = {P. A. P. Moran},
 journal = {Biometrika},
 number = {1/2},
 pages = {17-23},
 title = {Notes on Continuous Stochastic Phenomena},
 volume = {37},
 year = {1950}
}

@article{Ng2006,
 author = {Serena Ng},
 journal = {Journal of Business \& Economic Statistics},
 number = {1},
 pages = {12-23},
 title = {Testing Cross-Section Correlation in Panel Data Using Spacings},
 volume = {24},
 year = {2006}}

@article{Pesaran2021,
 author = {Pesaran, M. Hashem},
 journal = {Empirical Economics},
 number = {1},
 pages = {13--50},
 title = {General diagnostic tests for cross-sectional dependence in panels},
 volume = {60},
 year = {2021}}

@article{PUY2008,
 author = {Pesaran, M. Hashem and Ullah, Aman and Yamagata, Takashi},
 journal = {The Econometrics Journal},
 number = {1},
 pages = {105-127},
 title = {A bias-adjusted \uppercase{LM} test of error cross-section independence},
 volume = {11},
 year = {2008}}

@article{PY21,
title = {Estimation and inference in spatial models with dominant units},
journal = {Journal of Econometrics},
volume = {221},
number = {2},
pages = {591-615},
year = {2021},
author = {M. Hashem Pesaran and Cynthia Fan Yang},
}

@unpublished{Rose2018,
 author = {Christiern Rose},
 title = {Identification of Spillover Effects using Panel Data},
 note = {Working paper},
 year = {2018}}

@article{SYR2009,
 author = {Vasilis Sarafidis and Takashi Yamagata and Donald Robertson},
 journal = {Journal of Econometrics},
 number = {2},
 pages = {149-161},
 title = {A test of cross section dependence for a linear dynamic panel model with regressors},
 volume = {148},
 year = {2009}}

@article{SW2008,
 author = {Stock, James H. and Watson, Mark W.},
 journal = {Econometrica},
 number = {1},
 pages = {155--174},
 title = {Heteroskedasticity-Robust Standard Errors for Fixed Effects Panel Data Regression},
 volume = {76},
 year = {2008}}

\end{document}